\documentclass[useAMS,usenatbib]{mn2e}

\usepackage{graphics,graphicx}
\usepackage{paralist}
\usepackage{amsmath}
\usepackage{amssymb}
\usepackage{lscape}
\usepackage{rotating}
\usepackage{afterpage}

\title[Hot ISM in star-forming galaxies]{X-ray emission from star-forming galaxies -- II.  Hot interstellar medium}

\author[S. Mineo, M. Gilfanov and R. Sunyaev]{S. Mineo$^{1,2}$\thanks{E-mail:
smineo@head.cfa.harvard.edu}, M. Gilfanov $^{2, 3}$ and R. Sunyaev $^{2, 3}$\\
$^{1}$Harvard-Smithsonian Center for Astrophysics, 60 Garden Street Cambridge, MA 02138, USA\\
$^{2}$Max Planck Institut f\"ur Astrophysik, Karl-Schwarzschild-Str. 1 85741 Garching, Germany\\
$^{3}$Space Research Institute of Russian Academy of Sciences, Profsoyuznaya 84/32, 117997 Moscow, Russia}
\begin{document}
\sloppypar

\date{Accepted. Received; in original form}

\pagerange{\pageref{firstpage}--\pageref{lastpage}} \pubyear{2012}

\maketitle

\label{firstpage}

\begin{abstract}
We study the emission from the hot interstellar medium in a sample of nearby late type galaxies defined in Paper I. Our sample covers a broad range of star formation rates, from $\sim 0.1~M_\odot$/yr to   $\sim 17 ~M_\odot$/yr and stellar masses, from $\sim3\cdot 10^{8}~M_\odot$ to $\sim 6\cdot 10^{10}~M_\odot$. We take special care of systematic effects and contamination from bright and faint compact sources. We find that in all galaxies at least one optically thin thermal emission component is present in the unresolved emission, with the average temperature of $\left< kT\right>= 0.24$ keV. In about $\sim1/3$ of galaxies, a second, higher temperature component is required, with the $\left< kT\right>= 0.71$ keV. Although statistically significant variations in temperature between galaxies are present, we did not find any meaningful trends with the stellar mass or star formation rate of the host galaxy. The apparent  luminosity of the diffuse emission in the 0.5--2 keV band linearly correlates with the star formation rate with the scale factor of $L_{\rmn{X}}/\rmn{SFR}\approx 8.3\cdot 10^{38}$ erg/s per $M_\odot$/yr, of which in average $\sim 30-40\%$ is likely produced by faint compact sources of various types. We attempt to estimate the bolometric luminosity of the gas and and obtained results differing by an order of magnitude, $\log(L_{\rmn{bol}}/\rmn{SFR})\sim 39-40$, depending on whether intrinsic absorption in star-forming galaxies was allowed or not. Our theoretically most accurate, but in practice the most model dependent result for the {\em intrinsic} bolometric luminosity of ISM is $L_{\rmn{bol}}/\rmn{SFR}\sim 1.5\cdot 10^{40}$ erg/s per $M_\odot$/yr. Assuming that core collapse supernovae are the main source of energy,  it implies that $\epsilon_{SN}\sim 5\cdot 10^{-2} \left(E_{SN}/10^{51}\right)^{-1}$ of mechanical energy of supernovae  is converted into thermal energy of  ISM.
\end{abstract}


\begin{figure*}
\begin{center}
\hbox
{
\includegraphics[width=59mm]{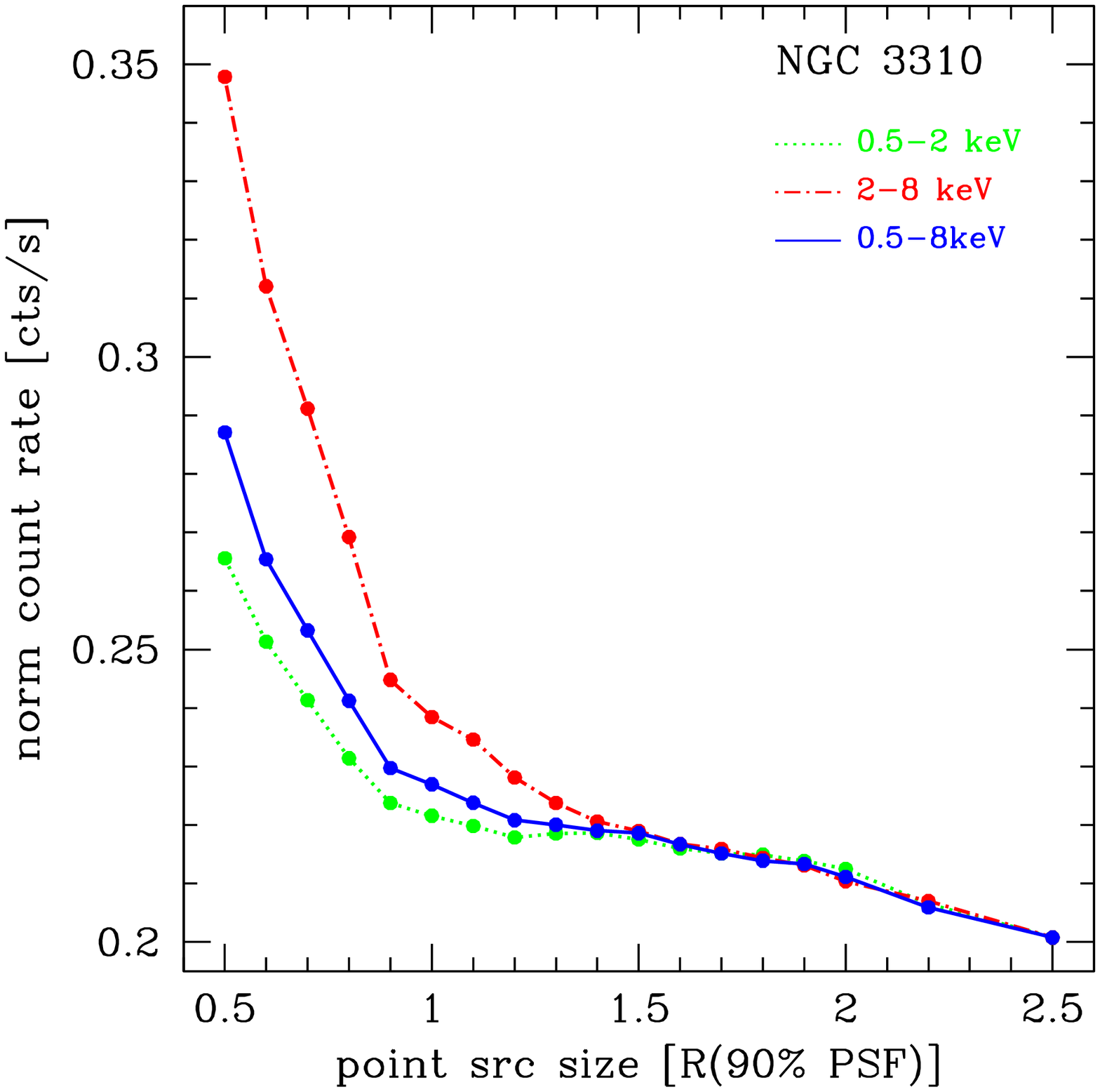}
\includegraphics[width=59mm]{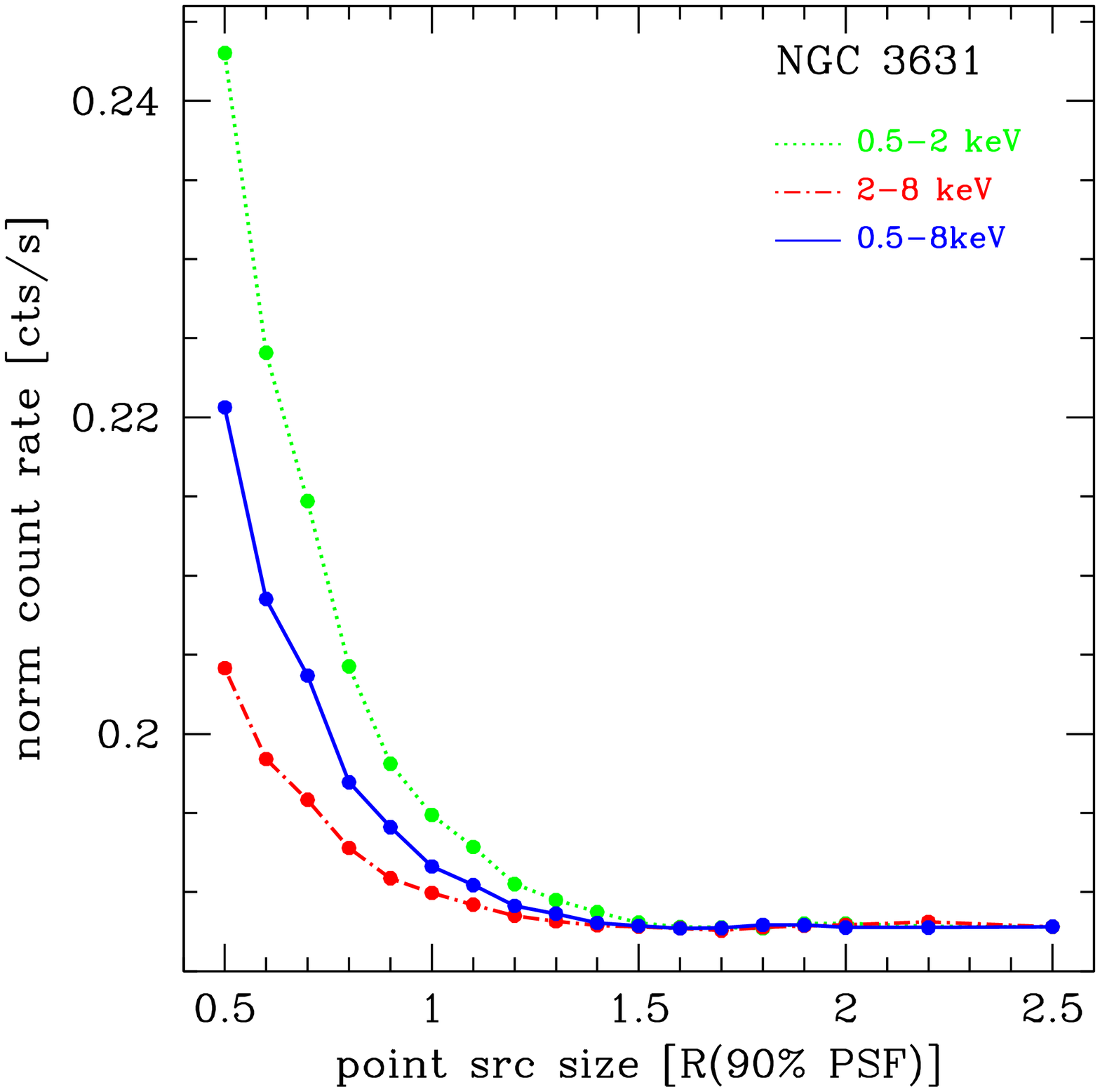}
\includegraphics[width=59mm]{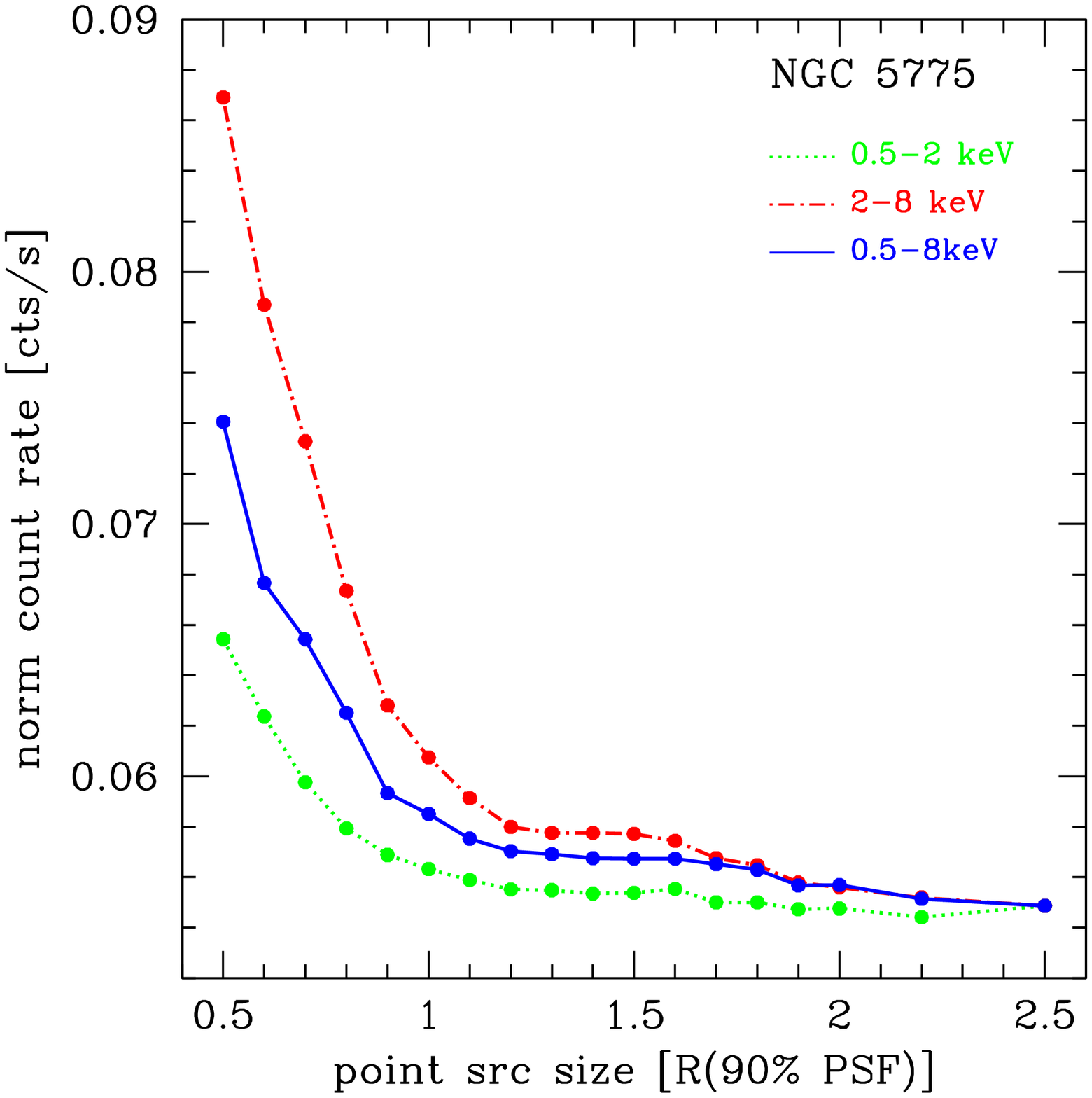}
}
\caption{Count rate of diffuse emission after the source removal versus the radius of the source region used for the point source removal, for the three spiral galaxies chosen as representative of our sample: NGC~3310, NGC~3631 and NGC~5775. The curves in each panel  are normalized to give the same count rate at a source cell size of 2.5. The dotted line (green) indicates the count rate in 0.5--2 keV band, the dash-dotted line (red) represents the count rates in 2--8 keV band, the dot solid curve (blue) shows the count rate in 0.5--8 keV band. The source region radius is expressed in in units of $R_{90\%\,\rmn{PSF}}$ (see Sect. \ref{sec:ps_removal} for details). Note that NGC 3310 has the smallest, by a factor of $\sim 1.6$, value of the $D25$ major radius.}
\label{fig:flux_r}
\end{center}
\end{figure*}

\section{Introduction}

It is well established that the number of high-mass X-ray binaries and their collective luminosity scale with the star formation rate (SFR) of the host galaxy \citep[Paper I hereafter]{2003MNRAS.339..793G, 2010ApJ...724..559L, 2012MNRAS.419.2095M}. This fact is well understood in terms of the short evolutionary time scales of high-mass X-ray binaries (HMXBs), both theoretically (e.g. Verbunt \& van den Heuvel 1995) and observationally \citep[e.g.][]{2007AstL...33..437S}.

Star-forming galaxies are also known to possess significant amounts of hot ionized gas of $\sim$sub-keV temperatures, which is a source of copious X-ray emission \citep[e.g.][]{2005ApJ...628..187G, 2006A&A...448...43T, 2009MNRAS.394.1741O, 2012arXiv1201.0551L}. The morphology of the diffuse X-ray emission suggests \citep[e.g.][]{2000AJ....120.2965S} that the gas is in the state of outflow,  driven by the collective effect of supernovae and winds from massive stars \citep[e.g.][]{1985Natur.317...44C}.
\citet{2004ApJ...610..213T} demonstrated  that the spatial distribution  of soft diffuse X-ray emission correlates with the sites of recent star formation in spiral arms as  traced by mid-infrared and H$_{\alpha}$ emission.  The total gas luminosity was found to generally correlate with  the star formation rate of the  host galaxy \citep[e.g.][]{2005ApJ...628..187G, 2009MNRAS.394.1741O,2012arXiv1201.0551L}. Although the majority (if not all) of the studies agree on the presence of such correlation, its slope and the scale factors determined by different authors  differ considerably. One part of this discrepancy could in principle be caused by the difference in the galaxy samples (e.g. ULIRGs versus more normal spiral galaxies) or analyzed regions (e.g. halos versus disks). But at least some part of the discrepancy  is caused by the difference in the data analysis methods, in particular in the way the contribution of compact sources was separated from the truly diffuse emission.  

The goal of this paper is to study the properties of the hot interstellar medium (ISM) in the sample of nearby star-forming galaxies from Paper I, with the main emphasis on the relation between X-ray luminosity of gas  and star formation rate of the host galaxy. We aim to notably improve the previous studies in terms of accuracy of the treatment of systematic effects (such as spill-over of counts from bright sources) as well as in terms of numbers of galaxies for which uniform $L_{\rmn{X}}-\rmn{SFR}$ data is available.

The structure of the paper is as follows. In Section 2 we briefly describe our sample. In Section 3 we present the techniques used to isolate the hot ISM and obtain its X-ray spectra and in Section 4 we consider the remaining contribution of faint discrete sources to the unresolved X-ray emission. In Section 5 we present the results of  the spectral analysis of diffuse emission. The X-ray luminosity of the ISM emission is obtained in Section 6 and its relation with the SFR of the host galaxy is investigated. In Section 7 we discuss our results.

\section{Sample selection and data preparation}
\label{sec:resolved_sample}

The goal of the present work is to study the unresolved X-ray emission from star-forming galaxies. To achieve this goal the latter must be separated from the emission of compact X-ray sources. We constructed a sample of galaxies based on the "primary sample" defined in Sect. 2.1 of Paper I. We selected galaxies whose full extent lies within a single chip of the {\it Chandra} detector. As the background correction is normally performed on a chip-to-chip basis, the latter criterion allows a straightforward and accurate background level estimation and subtraction, addressing both the instrumental and cosmic X-ray background (CXB). We selected and analyzed only single observations having exposure-time longer than $20 \rmn{ks}$. The thus constructed sample includes 21 galaxies (Table \ref{table:spectral_analysis}) whose star formation rates span a range from $\sim 0.1$ to $\sim 20$ $M_{\odot}$ yr$^{-1}$, similar to that of the primary sample of Paper I. 

The data preparation was done following the standard CIAO\footnote{http://cxc.harvard.edu/ciao3.4/index.html} threads (CIAO version 3.4; CALDB version 3.4.1), exactly in the same way as described in Sect. 3 of Paper I.

\section{Isolating the emission from the hot ISM}
\label{sec:ism}

The truly diffuse emission from the hot interstellar medium is polluted by several contaminating factors. These factors are: (i)  "spill-over" counts from bright resolved compact sources that have been removed from the image; (ii) unresolved faint compact X-ray sources; (iii) instrumental and cosmic X-ray backgrounds. In order to estimate the X-ray emission from hot ISM, the contribution of these components needs to be computed and removed.  In this and the next section we describe the procedures used to isolate the truly diffuse emission from contaminating components, to obtain the ISM spectrum and calculate its luminosity.

\subsection{Compact X-ray sources and their residual counts}
\label{sec:ps_removal}

The point sources analysis procedures used in Paper I were tuned to produce a sample with optimal point source detection sensitivity,  controlled incompleteness and accurate point source photometry. In contrast to that, the point source handling procedures used in this study are aimed to minimize the contamination from spill-over counts from bright sources, to minimize contribution of unresolved emission from  faint compact sources and to facilitate its accurate estimation for further correction.   

With this in mind, we start with the determination of the optimal size of the region to be used to remove the point sources counts from the image.  To this end we selected three spiral galaxies, representative of our sample: a face-on galaxy with strong diffuse emission (NGC~3310), a face-on galaxy with weak diffuse emission (NGC~3631) and a edge-on galaxy (NGC~5775). To each galaxy,  the following procedure was applied. We used the wavelet-based source detection algorithm {\tt wavdetect} (see Sect. 3.1 of Paper I for details on the parameter settings) to search for point like sources in the soft ($0.5-2.0$ keV), hard ($2.0-8.0$ keV) and total ($0.5-8.0$ keV) energy bands.  For each source, we used the information about the shape of the point spread function (PSF) in the source position to determine the radius of the region containing 90\% of source counts, following the method described in Paper I. In the following, we refer to this radius as $R_{90\%\,\rmn{PSF}}$. 

Based on the source lists obtained in each energy band, we created a set of source regions having radii ranging from $0.5\cdot R_{90\%\,\rmn{PSF}}$ to $2.5\cdot  R_{90\%\,\rmn{PSF}}$ with a step of 0.1. A corresponding set of diffuse emission images was created for each galaxy, in each energy band, adopting the following method. Using the CIAO task {\tt dmfilth}, we removed the source regions from the image and filled in the holes left by the source removal with pixel values interpolated from surrounding background regions (POISSON method). For the background region we used a circle with radius 3 times the radius of the source region. In order to avoid biases in the interpolation, we ensured that the chosen background annuli did not contain neighboring point sources. For each background region listed in the input file, we subtracted all the overlapping neighboring point source regions and merged them into a single source removal region. 

For each such image we estimated the count rate within the $D25$ ellipse  using the CIAO task {\tt dmextract}. The dependence of the count rate on the radius of the source region is shown in Fig. \ref{fig:flux_r}. 
In all three galaxies, the count rate appears to follow the same trend -- a rather sharp decrease at small source radii  $\leq R_{90\%\,\rmn{PSF}}$ is followed by a much slower trend and the curves become nearly flat at $R>1.5 R_{90\%\,\rmn{PSF}}$. Some residual decrease beyond $R\sim1.5 R_{90\%\,\rmn{PSF}}$ is observed in the case of NGC 3310, which is likely caused by the fact that at large values of $R$  the source region size becomes comparable with the characteristic angular scale of the surface brightness variations. Note that this galaxy has the smallest, by a factor of 1.6, value of $D25$ major radius among the three galaxies selected for this analysis. In any case, the remaining variations of the diffuse emission count rate do not exceed $\approx 10\%$ level even for this galaxy.

Based on this analysis we adopted the source region radius of  $R=1.5 R_{90\%\,\rmn{PSF}}$. This value is a reasonable compromise, minimizing the contamination of diffuse emission by point source counts without compromising the statistics for the diffuse emission itself.

\subsection{Spillover counts from ULXs}
\label{sec:ulx_spill_counts}

\begin{figure}
\begin{center}
\includegraphics[width=1.0\linewidth]{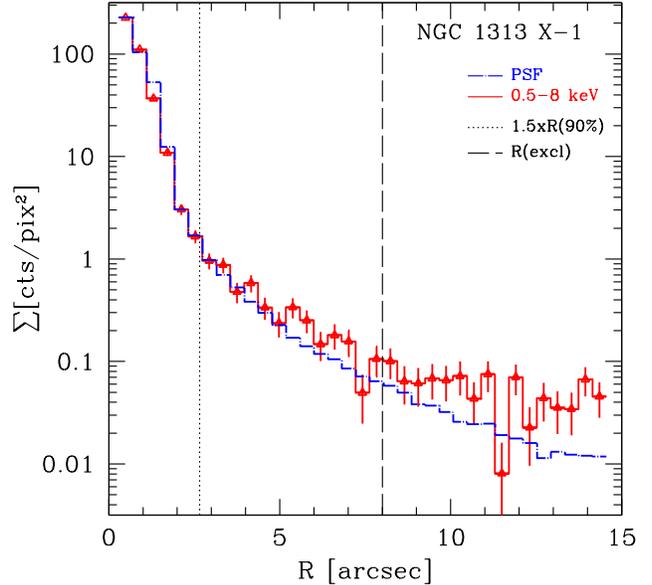}
\caption{An example of the analysis performed to minimize the contamination of the diffuse emission by  spillover counts from bright point sources. The figure shows the full band ($0.5-8$ keV) surface brightness profile centered at NGC1313  X-1 (solid line, red) and the radial profile of the {\it Chandra} ACIS-S PSF at  this position (dash-dotted line, blue). The vertical dotted line represents the default source region radius of $1.5 R_{90\%\,\rmn{PSF}}$, the vertical dashed line is the $8\arcsec$ radius used for bright point sources (see Sect. \ref{sec:ulx_spill_counts} for details).}
\label{fig:ulx_sb}
\end{center}
\end{figure}

The procedure to optimize the source region size described in the previous sub-section  allowed us to minimize the contamination of diffuse emission by spillover counts from the majority of point sources. However, the  residual counts from extremely bright compact sources may still pollute  the unresolved X-ray emission. After visual inspection, in 5 galaxies (NGC~1313, NGC~4214, NGC~4490, NGC~5474 and NGC~7793) we detected the presence of an excess of X-ray counts around very bright compact sources after their removal. For these sources, the counts located outside the default source region can make up to $\sim10\%$ of the total count rate of the unresolved emission.  In these cases the size of the source region needs to be increased. 

In order to determine the required source region size, we extracted the {\it Chandra} PSF in the positions of bright point sources using the {\tt mkpsf} task. The surface brightness profile centered on the positions of bright sources was constructed for both the observed counts in 0.5--8 keV band and the PSF. An example of such profiles for NGC1313 X-1 is shown in Fig. \ref{fig:ulx_sb}. As it is clear from the figure, the measured profile follows the PSF up to at least $\sim5-8\,\rmn{arcsec}$, indicating that the surface brightness is dominated by counts for the bright point source and not by the diffuse emission itself. Therefore, excluding a circle with radius $R=1.5\cdot R_{90\%\,\rmn{PSF}}$ for these  sources is insufficient.  For these sources we decided to use the source region radii in the $\sim8-10\,\rmn{arcsec}$ range, depending on the brightness of the source.

\begin{figure*}
\begin{center}
\hbox
{
\includegraphics[width=59mm]{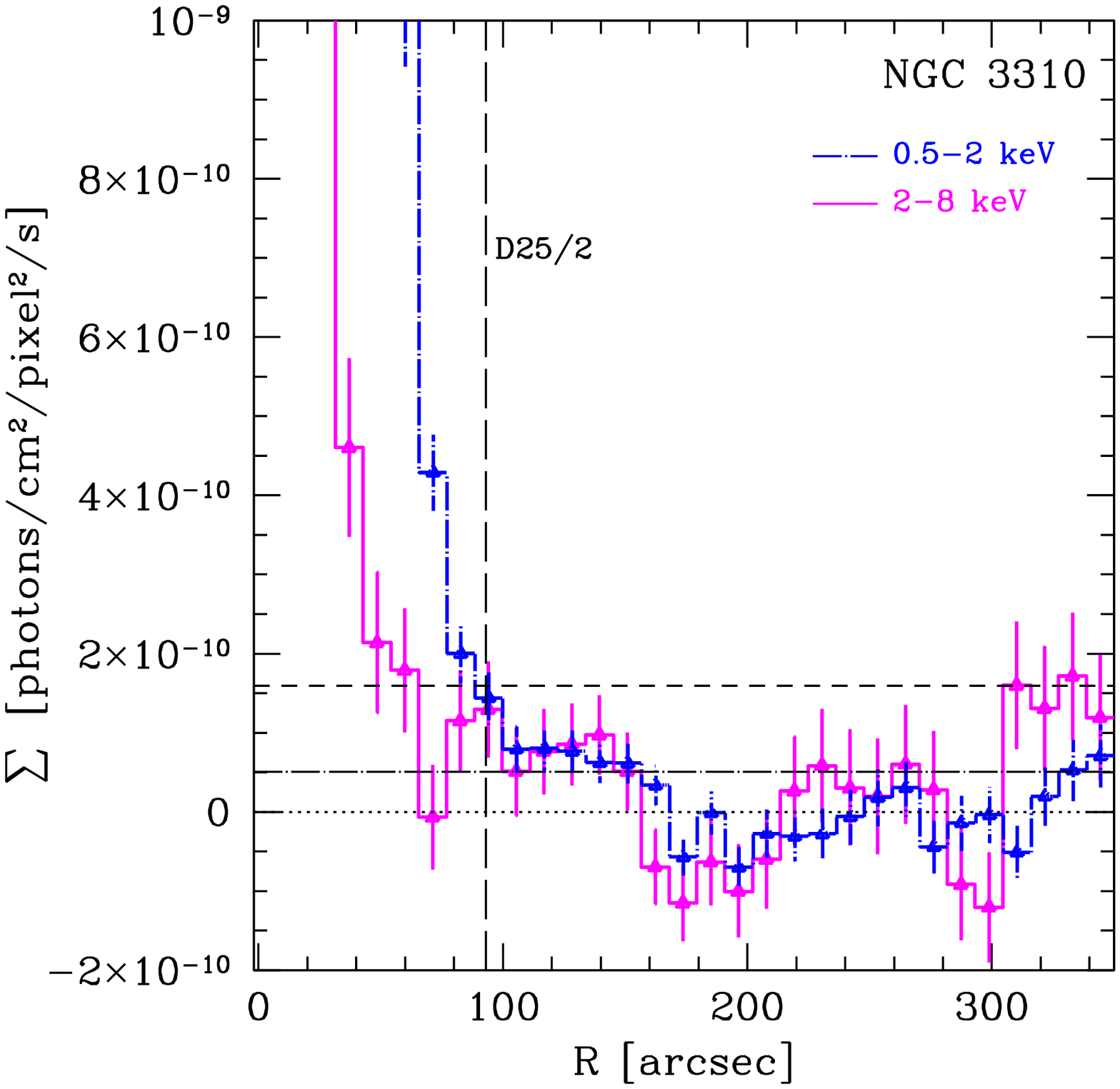}
\includegraphics[width=59mm]{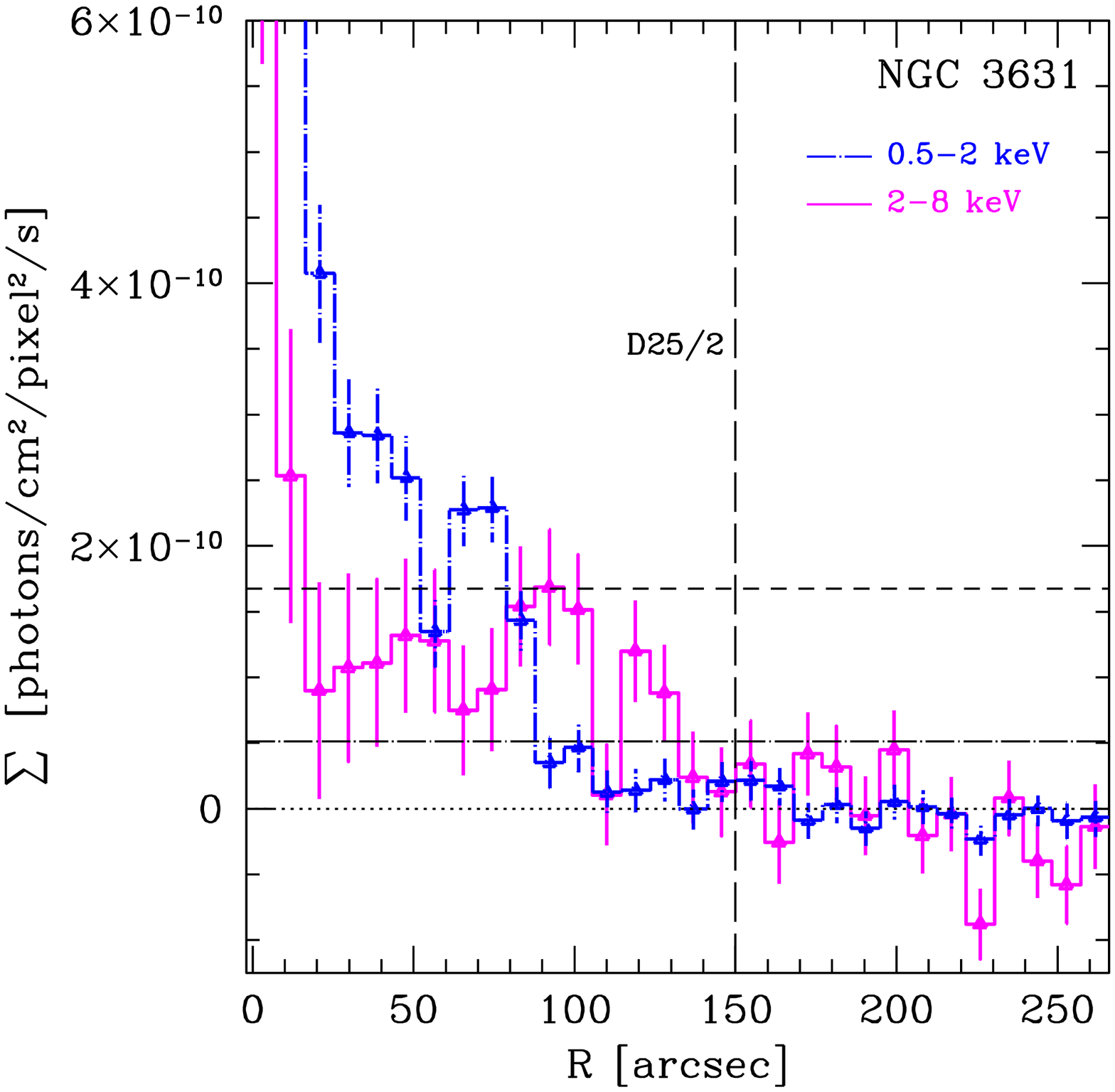}
\includegraphics[width=59mm]{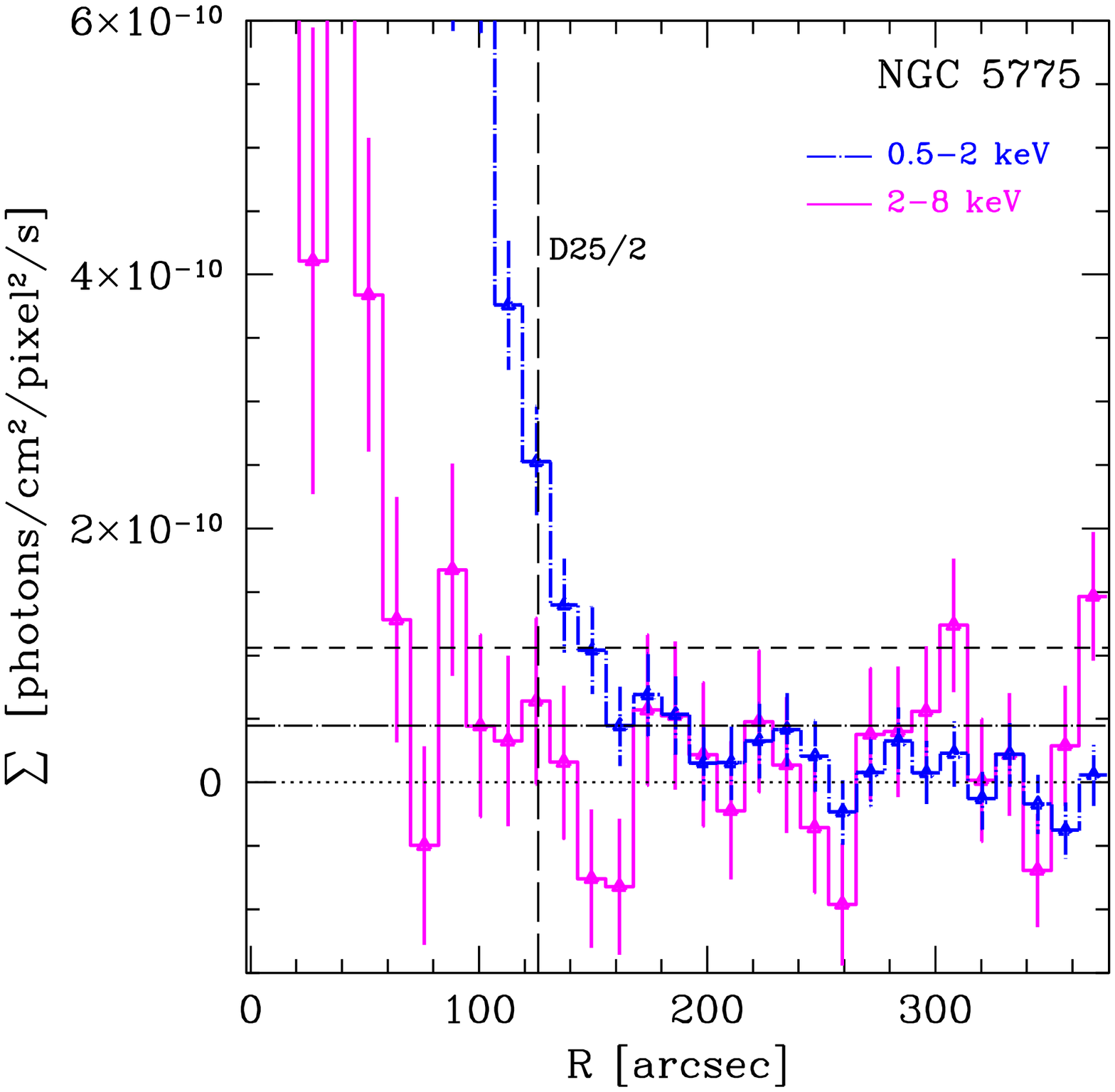}
}
\caption{An example of exposure-corrected background-subtracted surface brightness profiles in the soft (dash-dotted line, blue) hard (solid line, magenta) bands for 3 representative galaxies. The statistically significant brightness fluctuations around zero level are present in some case. These variations are stronger in the hard band,  can have both positive and negative amplitude and do not exceed 10\% of the background level. This level is indicated by horizontal thin lines for the soft (dash-dotted horizontal line) and hard (dashed horizontal line) energy bands. This level characterizes the accuracy of our background subtraction procedure. The vertical dashed line indicates the radius corresponding to the $D25$ region.}
\label{fig:sb_profile_roundzero}
\end{center}
\end{figure*}

\subsection{Instrumental and cosmic X-ray background}
\label{sec:bkg_removal}

The instrumental and cosmic X-ray background was determined from the background regions located on the same chip, outside the body of the galaxy. In order to determine the exact locations of the background regions, for each galaxy we constructed exposure-corrected radial profiles, with no background subtraction, in the soft (0.5--2 keV) and hard (2--8 keV) energy bands. The surface brightness profiles were  constructed  in concentrical elliptical annuli, around the center of the galaxy, with shape parameters --  axis ratio and the position angle, set equal to the corresponding parameters of the optical emission \citep{1991trcb.book.....D}. The profiles were extracted over the whole {\it Chandra} detector area.  All profiles showed clear flattening in the outer parts, indicating that  no significant X-ray emission from the target galaxy was present there. The background  region for each galaxy was defined  as the part of the Chandra CCD outside the  $D25$, $1.5\cdot D25$ or $2\cdot D25$ region, depending on the shape of the  emission profile of the galaxy.

Once the background regions were defined, we constructed for each galaxy the exposure-corrected background-subtracted surface brightness profiles, in order to check the accuracy  of the background determination procedure. Examples of such profiles are shown in Fig. \ref{fig:sb_profile_roundzero}, for the same 3 galaxies as used for the source region analysis. We found that in some  galaxies statistically significant surface brightness variations are present at large radii, where no emission from the galaxy itself is expected. 

These variations are typically stronger in the hard band. When present, they are equally strong in the positive and negative amplitude, suggesting that they may be resulting from systematic effects. However, as the zero level was determined from the data itself, their physical origin can not be excluded either. Their amplitude is within $10\%$ of the background level and, to be conservative, we accepted this as a threshold in analyzing the spatial distribution of the diffuse emission. For all galaxies, this threshold was reached near the $D25$ boundary of the galaxy.

\section{Unresolved emission and faint compact sources}
\label{sec:contamination}

\subsection{Spectra of unresolved emission}
\label{sec:l95}

In order to obtain the unresolved emission spectra with well controlled contribution of unresolved faint point sources we defined the luminosity limit  $L_{\rmn{95}}$ corresponding to a rather high level of the source detection completeness -- 95\%. Point sources with luminosity above this limit were excluded using the source regions determined in Sect. \ref{sec:ps_removal} and \ref{sec:ulx_spill_counts}. The sources below this limit, even though statistically significantly detected by the {\tt wavdetect} task,  were left on the image. This method allows us to accurately estimate the contribution of various types of faint compact sources to the unresolved emission.   

The spectra of the point source-free emission were extracted using the CIAO {\tt specextract} script. To be consistent with the analysis performed in Paper I,  the extraction regions were  same as defined in the Section 4 of Paper I, ranging from $\sim D_{25}$ for moderately sized galaxies to $\sim 3/4-1/2~D_{25}$ for the largest ones, the bulges being excluded when possible. We grouped the spectra in order to have minimum 15 counts per channel to apply the $\chi^2$ fitting. The nominal background spectra were extracted as described in the Sect. \ref{sec:bkg_removal}. The resulting background subtracted spectra of unresolved emission are shown in Fig. \ref{fig:diffuse_spectra}.

\subsection{Unresolved high-mass X-ray binaries}
\label{sec:hmxb_removal}

\begin{table}
\centering
\begin{minipage}{80mm}
\caption{Completeness limits and luminosities of unresolved HMXBs.}
\label{table:unres_hmxb}
\begin{tabular}{@{}l c l c @{}}
\hline
& & \multicolumn{2}{|c|}{{\sc unresolved HMXBs}\,\footnote{Soft and hard band luminosity of unresolved HMXBs computed as described in the Section \ref{sec:hmxb_removal}.}} \\
Galaxy & $L_{\rmn{95}}$\footnote{The luminosity threshold used for removal of compact sources. It is  defined as the 0.5--8 keV luminosity corresponding to 95\% source detection efficiency (see sect. \ref{sec:l95}).} &\vline\, $L_{0.5-2\,\rmn{keV}}$ & $L_{2-8\,\rmn{keV}}$ \\
 & $(\rmn{erg}\,\rmn{s}^{-1})$ &\vline\, $(\rmn{erg}\,\rmn{s}^{-1})$ & $(\rmn{erg}\,\rmn{s}^{-1})$ \\	
\hline
\hline
NGC  0278 & $2.13\times 10^{38}$ &\vline\, $8.35\times 10^{37}$ & $1.74\times 10^{38}$ \\ 
NGC 0520 & $1.96\times 10^{38}$ &\vline\, $7.55\times 10^{38}$ & $1.57\times 10^{39}$ \\ 
NGC 1313 & $8.22\times 10^{36}$ &\vline\, $2.11\times 10^{37}$ & $4.39\times 10^{37}$ \\ 
NGC 1569 & $1.06\times 10^{36}$ &\vline\, $1.43\times 10^{36}$ & $2.98\times 10^{36}$ \\ 
NGC 2139 & $1.84\times 10^{38}$ &\vline\, $1.16\times 10^{39}$ & $2.42\times 10^{39}$ \\ 
NGC 3079 & $2.26\times 10^{38}$ &\vline\, $7.31\times 10^{38}$ & $1.52\times 10^{39}$ \\ 
NGC 3310 & $1.53\times 10^{38}$ &\vline\, $1.50\times 10^{39}$ & $3.12\times 10^{39}$ \\ 
NGC 3556 & $2.97\times 10^{37}$ &\vline\, $2.82\times 10^{38}$ & $5.86\times 10^{38}$ \\ 
NGC 3631 & $1.12\times 10^{38}$ &\vline\, $1.03\times 10^{39}$ & $2.15\times 10^{39}$ \\ 
NGC 4038/39 & $1.53\times 10^{38}$ &\vline\, $9.55\times 10^{38}$ & $1.99\times 10^{39}$ \\ 
NGC 4194 & $3.79\times 10^{39}$ &\vline\, $5.09\times 10^{39}$ & $1.06\times 10^{40}$ \\ 
NGC 4214 & $1.81\times 10^{37}$ &\vline\, $1.98\times 10^{37}$ & $4.12\times 10^{37}$ \\ 
NGC 4490 & $1.41\times 10^{38}$ &\vline\, $6.64\times 10^{38}$ & $1.38\times 10^{39}$ \\ 
NGC 4625 & $1.73\times 10^{37}$ &\vline\, $4.54\times 10^{36}$ & $9.45\times 10^{36}$ \\ 
NGC 5253 & $2.22\times 10^{37}$ &\vline\, $4.14\times 10^{36}$ & $8.62\times 10^{36}$ \\ 
NGC 5474 & $1.30\times 10^{37}$ &\vline\, $4.34\times 10^{37}$ & $9.04\times 10^{37}$ \\ 
NGC 5775 & $1.41\times 10^{38}$ &\vline\, $1.52\times 10^{39}$ & $3.16\times 10^{39}$ \\ 
NGC 7090 & $8.55\times 10^{37}$ &\vline\, $1.44\times 10^{38}$ & $2.99\times 10^{38}$ \\ 
NGC 7541 & $4.22\times 10^{38}$ &\vline\, $1.98\times 10^{39}$ & $4.12\times 10^{39}$ \\ 
NGC 7793 & $3.37\times 10^{37}$ &\vline\, $7.99\times 10^{36}$ & $1.66\times 10^{37}$ \\ 
UGC 05720 & $6.25\times 10^{38}$ &\vline\, $6.15\times 10^{38}$ & $1.28\times 10^{39}$ \\ 
\hline
\end{tabular}
\end{minipage}
\end{table}

The contribution of unresolved HMXBs to the X-ray diffuse emission spectrum of each galaxy was estimated in the statistical way.

\begin{figure}
\begin{center}
\includegraphics[width=0.7\linewidth, angle=270]{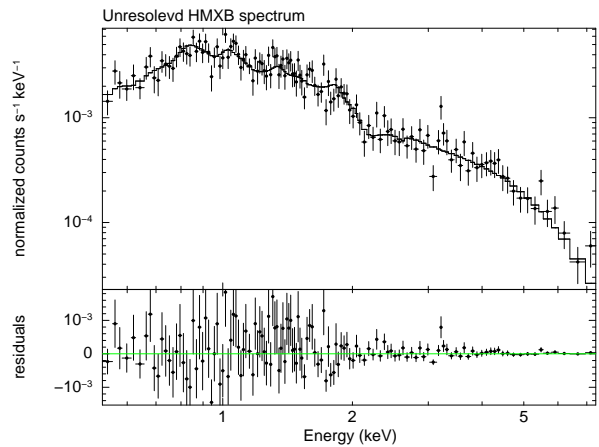}
\caption{The average spectrum of unresolved high-mass X-ray binaries. It was obtained by combining the spectra of all the compact sources with $\log L_{\rmn{X}} < 37.5$ detected in the resolved galaxies from our sample (see Sect. \ref{sec:hmxb_removal} for details). The best-fitting model includes two thermal components with temperature of $0.13$ keV and $0.45$ keV respectively and a power law with slope $1.8$, absorbed by a hydrogen column density of $N_{\rmn{H}} =  4.9 \cdot 10^{21}$ cm$^{-2}$.} 
\label{fig:unres_hmxb_spectra}
\end{center}
\end{figure}

In order to determine the spectrum of faint HMXBs we combined the spectra of all resolved point sources with $\log L_{\rmn{X}} < 37.5$ detected in the galaxies of our sample. The chosen luminosity limit allows us to include compact sources that are as faint as possible without compromising the statistics of the final combined spectrum. The spectrum of each detected point source was extracted from a circular region centered on the source position given by {\tt wavdetect}, the  radius of the source region was fixed at  $R_{90\%\,\rmn{PSF}}$. For the background region we used an annulus  with the inner and outer radius equal to $R_{90\%\,\rmn{PSF}}$ and $3\cdot R_{90\%\,\rmn{PSF}}$ respectively.  From this annulus  we subtracted all the overlapping neighboring point source regions. 
Average spectra of faint point sources in each galaxy were extracted and weighted ARFs and RMFs were obtained, using the tasks \texttt{dmextract}, \texttt{mkwarf} and \texttt{mkrmf}. These spectra were then further combined into one using the \texttt{combine\_spectra} task. The resulting spectrum, grouped to have minimum 15 counts per cannel is shown in Fig. \ref{fig:unres_hmxb_spectra}.

This spectrum was initially  modeled  with an absorbed power law model, however,  we noted the presence of excess counts at low energies. 
We modeled the soft excess with two thermal components with best fit temperatures of $0.13$ keV and $0.45$ keV. The other parameters of the fit were:  the power law slope $\Gamma=1.8$ and the hydrogen column density $N_{\rmn{H}} =  4.9 \cdot 10^{21}$ cm$^{-2}$, the reduced chi-squared $\chi^{2}_{\nu} = 1.009$ for 125 degrees of freedom. 

In order to compute the contribution of faint HMXBs, for each galaxy the spectrum was re-normalized  in order to give the expected $0.5-8$ keV band luminosity of unresolved HMXBs. The latter was calculated by integrating the average X-ray luminosity function (XLF) for HMXBs, eq. (18) of Paper I, between $10^{34}$ erg/s and the completeness limit $L_{95}$ (above which the detected compact sources were excluded). The XLF  normalization was adjusted for each galaxy to match the number of point sources detected above $L_{95}$(corrected for the expected contribution of background AGNs). The contribution of unresolved HMXBs is plotted for each galaxy  in Fig. \ref{fig:diffuse_spectra} and presented in a tabular form  in the Table \ref{table:unres_hmxb}, along with the  values of the completeness limit, $L_{\rmn{95}}$.

\begin{figure*}
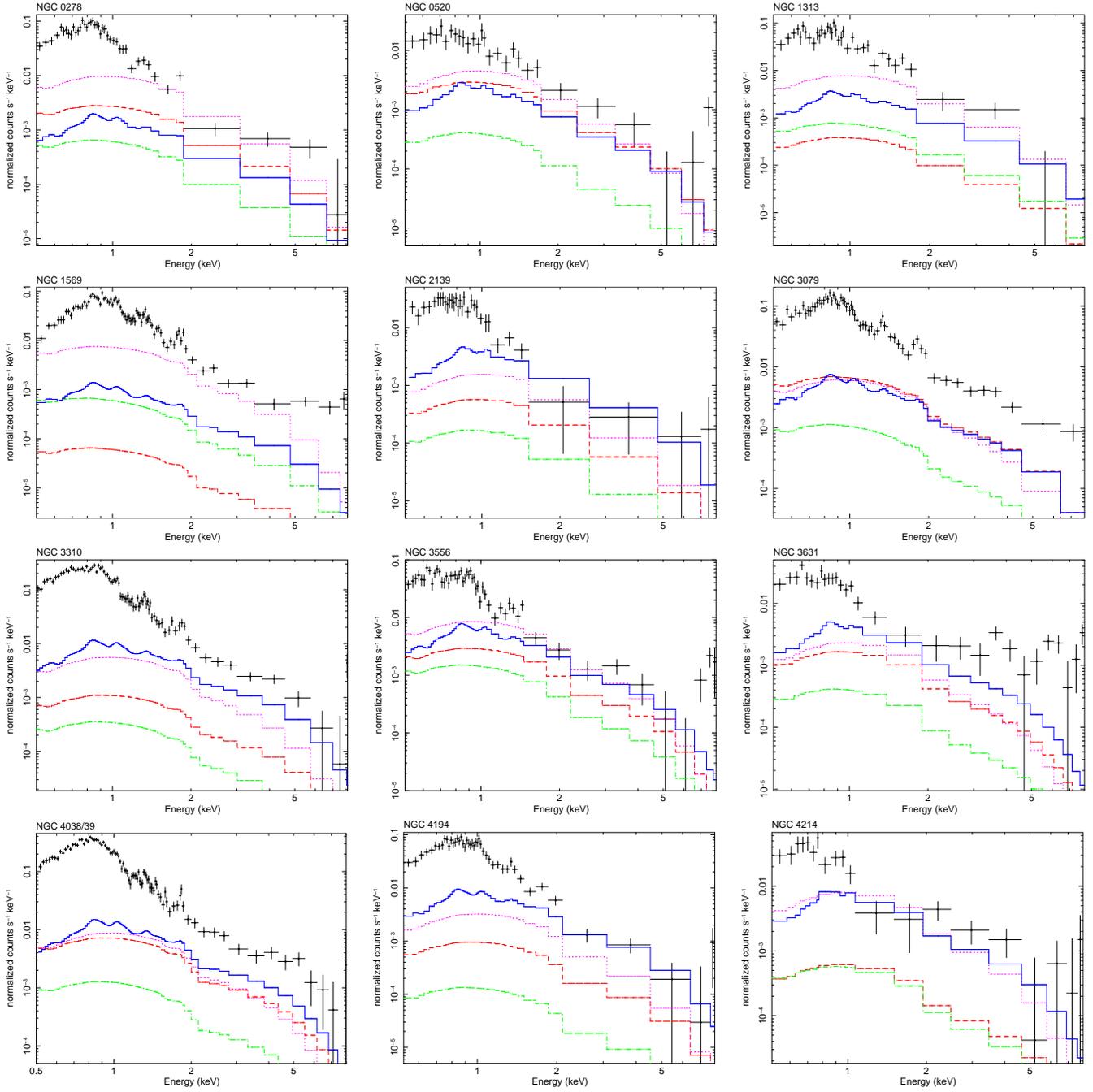

\centering
\mbox
{
\includegraphics[width=0.25\linewidth, angle=270]{ngc0278_unres_contribs.ps}                                                                                                      
\includegraphics[width=0.25\linewidth, angle=270]{ngc0520_unres_contribs.ps}   
\includegraphics[width=0.25\linewidth, angle=270]{ngc1313_unres_contribs.ps}                                                                                                      
}
\mbox
{                                                                                              
\includegraphics[width=0.25\linewidth, angle=270]{ngc1569_unres_contribs.ps}                                                                                                      
\includegraphics[width=0.25\linewidth, angle=270]{ngc2139_unres_contribs.ps}                                                                                                      
\includegraphics[width=0.25\linewidth, angle=270]{ngc3079_unres_contribs.ps}                                                                                                      
}
\mbox
{
\includegraphics[width=0.25\linewidth, angle=270]{ngc3310_unres_contribs.ps}                                                                                                      
\includegraphics[width=0.25\linewidth, angle=270]{ngc3556_unres_contribs.ps}                                                                                                      
\includegraphics[width=0.25\linewidth, angle=270]{ngc3631_unres_contribs.ps}                                                                                                      
}
\mbox
{
\includegraphics[width=0.25\linewidth, angle=270]{antennae_unres_contribs.ps}                                                                                                      
\includegraphics[width=0.25\linewidth, angle=270]{ngc4194_unres_contribs.ps}                                                                                                      
\includegraphics[width=0.25\linewidth, angle=270]{ngc4214_unres_contribs.ps}                                                                                                      
}
\caption{The observed spectra of diffuse X-ray emission for the sample of resolved galaxies. The models show the predicted contributions of unresolved HMXBs (solid curve, blue), LMXBs (dashed, red), CVs and ABs (dash-dotted, green), young stellar objects (dotted, magenta) estimated as described in Sect. \ref{sec:contamination}.} 
\label{fig:diffuse_spectra}
\end{figure*}

\addtocounter{figure}{-1}
\begin{figure*}
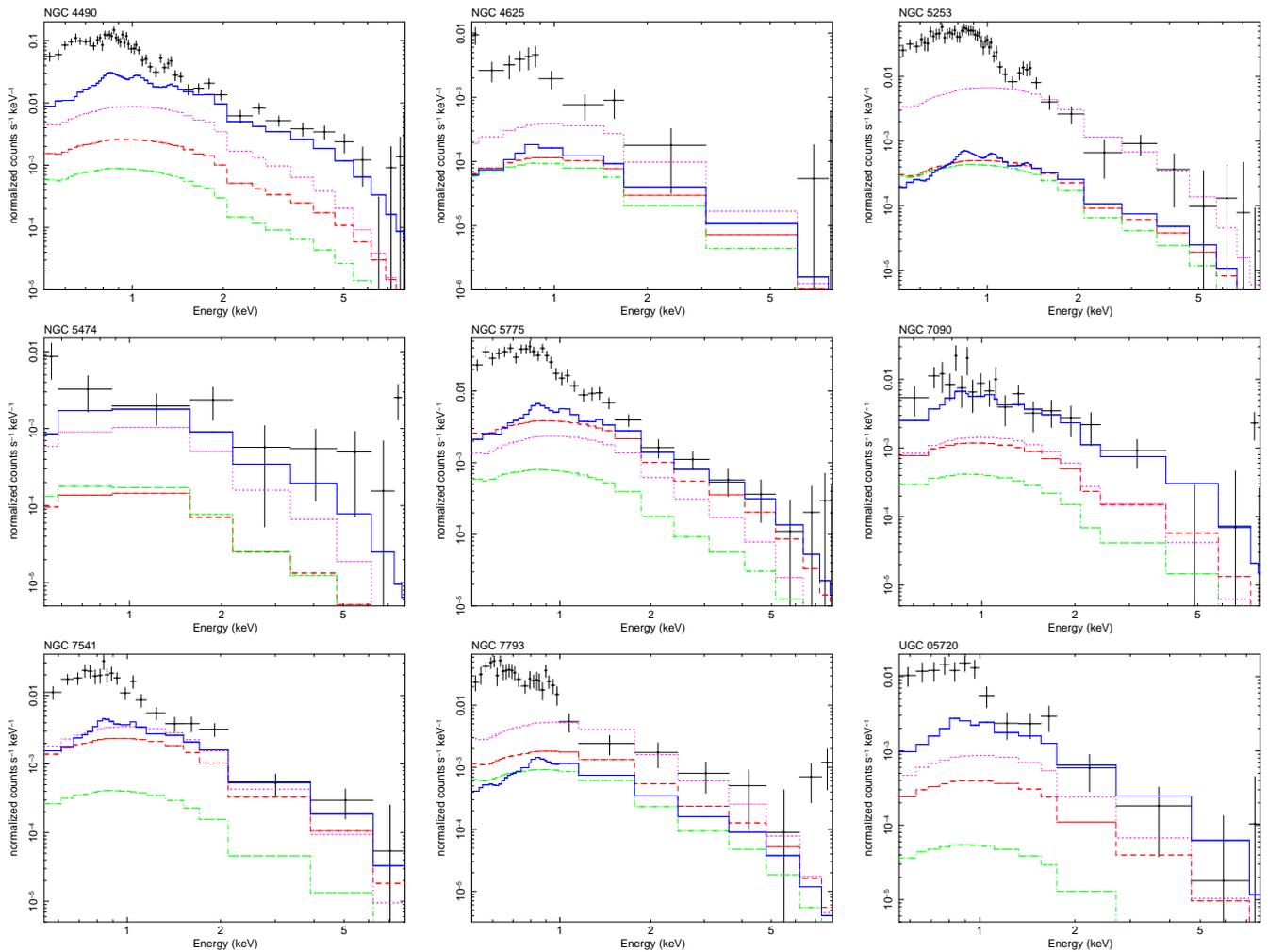

\centering
\mbox
{
\includegraphics[width=0.25\linewidth, angle=270]{ngc4490_unres_contribs.ps}                                                                                                      
\includegraphics[width=0.25\linewidth, angle=270]{ngc4625_unres_contribs.ps}                                                                                                      
\includegraphics[width=0.25\linewidth, angle=270]{ngc5253_unres_contribs.ps}  
}
\mbox
{
\includegraphics[width=0.25\linewidth, angle=270]{ngc5474_unres_contribs.ps}                                                                                                      
\includegraphics[width=0.25\linewidth, angle=270]{ngc5775_unres_contribs.ps}                                                                                                      
\includegraphics[width=0.25\linewidth, angle=270]{ngc7090_unres_contribs.ps}  
}
\mbox
{
\includegraphics[width=0.25\linewidth, angle=270]{ngc7541_unres_contribs.ps}                                                                                                      
\includegraphics[width=0.25\linewidth, angle=270]{ngc7793_unres_contribs.ps}                                                                                                      
\includegraphics[width=0.25\linewidth, angle=270]{ugc05720_unres_contribs.ps}  
}
\caption{Continued} 
\label{fig:diffuse_spectra}
\end{figure*}

\subsection{Unresolved LMXBs, CVs and ABs}
\label{sec:lmxb_removal}

In early type galaxies, the number and luminosity of low-mass X-ray binaries  (LMXBs) scales well with the stellar mass of the galaxy \citep{2004MNRAS.349..146G}. 

This may suggest that the contribution of unresolved LMXBs can be predicted from the stellar mass enclosed in the analyzed regions  of galaxies. However, as it was first noticed by \citet{2010A&A...512A..16B} and \citet{2012MNRAS.419.2095M}, the scale in the $L_{\rmn{X}}-M_{\star}$ relation for LMXBs appears to depend on the age of the stellar population, the $L_{\rmn{X}}/M_{\star}$ ratio being smaller in younger stellar environments. Therefore the average scaling relation of \citep{2004MNRAS.349..146G} will overestimate the contribution of unresolved LMXBs. This was further confirmed by \citet{2012arXiv1202.2331Z} who found strong age dependence of the $L_{\rmn{X}}/M_{\star}$ in 20 nearby early type galaxies. However, although \citet{2012arXiv1202.2331Z}  demonstrated the significance of the effect, they were not able to separate the age dependence from the dependence on the globular cluster content of the galaxy, and did not provide age-dependent scaling relations which could be used to predict numbers and luminosity of unresolved LMXBs.
On the other hand, as LMXBs are the (very) minor constituent of the compact source population in the analyzed regions, their scaling relation can not be renormalized for each galaxy individually, using the numbers of detected compact sources, in the same manner as the contribution of unresolved HMXBs was computed in the Section \ref{sec:hmxb_removal}.
These factors make the accurate calculation of the contribution of unresolved LMXBs to the unresolved emission  in star-forming galaxies  currently impossible.  On the other hand, one of the selection criteria in constructing the galaxy sample was designed to control the overall contribution of LMXBs (Section 2.1 of Paper I). The latter was further minimized by the way in which the analysis regions were selected (Section 4.2 of Paper I). For the above reasons, similar to Paper I,  we chose not to subtract the contribution of unresolved LMXBs.

For the same reason, although the $L_{\rmn{X}}-M_{\star}$ scaling relation is well established for cataclysmic variables (CVs) and active binaries (ABs) in Solar neighborhood \citep{2006A&A...450..117S} and in early type galaxies \citep{2011MNRAS.418.1901B}, it can not be directly applied to the stellar environment in  star-forming galaxies.  Following the same line of arguments, we chose not to subtract their contribution.  

In order to roughly access the  amplitude of the contribution of faint sources associated with the old stellar populations, we plot  in Fig. \ref{fig:diffuse_spectra}, along with the measured spectra of unresolved emission and with the contribution of unresolved HMXBs,  the spectra of unresolved LMXBs, CVs and ABs, estimated using the average scaling relations for these types of sources.  The LMXB spectrum was computed following the same method as used in the Section \ref{sec:hmxb_removal} for calculation of the contribution of unresolved HMXBs.  We used the average LMXB luminosity function and scaling relation from \citet{2004MNRAS.349..146G}, corresponding to $L_{\rmn{X}}(>10^{37} \rmn{erg/s})/M_{\odot}=8\times 10^{39}$ erg/s/$10^{11}M_{\odot}$. For the spectrum of faint LMXBs we assumed a power law with photon index $\Gamma=1.8$ and  $N_{\rmn{H}}=10^{21} \rmn{cm}^{-2}$. In computing the combined spectrum of CVs and ABs we used their average 2--10 keV scaling relation  for elliptical galaxies $L_{\rmn{X}}/L_K=3.1\cdot 10^{27}$ erg/s/$L_{K,\odot}$ determined by \citet{2011MNRAS.418.1901B}. For the effective average spectrum of CVs and ABs we assumed a flat power law spectrum ($\Gamma=2$), in agreement with the shape of the spectrum of unresolved emission in gas-poor elliptical galaxies \citep[e.g.][]{2010A&A...512A..16B}. To this spectrum we applied same absorption as for unresolved LMXBs,  $N_{\rmn{H}}=10^{21} \rmn{cm}^{-2}$. We emphasize that these calculations and, correspondingly,  the curves plotted in  Fig. \ref{fig:diffuse_spectra} represent an upper limit on the contribution of faint LMXBs. The same is possibly true for CVs and ABs.

\subsection{Unresolved young faint objects}
\label{sec:yso_removal}

Several types  of  intrinsically  faint X-ray objects are expected to be found in star-forming environments. These include protostars, young stellar objects, young stars and colliding wind binaries. Although they are intrinsically X-ray faint, their collective radiation can, in principle, contaminate the unresolved emission.  Unlike ellipticals, there are virtually no gas-free star-forming galaxies, therefore separation of faint objects from the truly diffuse ISM in the same manner as it was done for faint objects in early type galaxies \citep{2011MNRAS.418.1901B} is difficult, if not impossible. On the other hand, these objects typically have hard spectra, corresponding to temperatures of the order of a few keV, when approximated by bremsstrahlung emission model \citep{1999ARA&A..37..363F, 2007ApJ...669..493W}.  This is significantly larger than sub-keV temperatures, characteristic for the ISM in star-forming galaxies. With this in mind, \citet{2011MNRAS.418.1901B} assumed that the hard  unresolved emission of late type galaxies is due to radiation from faint compact objects and attempted to calibrate their scaling relation in the 2--10 keV energy band, obtaining $L_{\rmn{X}}/\rmn{SFR}\sim 1.7\cdot 10^{38}$ ($\rmn{erg}\,\rmn{s}^{-1})/(M_{\odot}\,\rmn{yr}^{-1})$, which is about $\sim 7\%$ of the total HMXB luminosity per SFR and comparable to the contribution of unresolved LMXBs in our sample. We use this relation to indicate the possible contribution of faint young compact sources to the unresolved emission in Fig.\ref{fig:diffuse_spectra} along with other components. In computing it we assumed that  their spectra can be described by bremsstrahlung emission with temperature $kT \sim 3$ keV \citep{2007ApJ...669..493W}, to which we applied same absorption as for unresolved LMXBs, CVs and ABs, $N_{\rmn{H}}=10^{21} \rmn{cm}^{-2}$.

\subsection{Contamination by faint compact sources -- Conclusions and caveats}
\label{sec:contam_concl}

Out of several different types of unresolved compact sources, the contribution of only one, faint HMXBs, can be accurately calculated and removed from the unresolved emission.
For other types of sources (LMXBs, CVs and ABs, young stellar objects),  observations provide some useful  constraints which can be used to approximately characterize their contribution. These predictions are summarized in Table \ref{table:unres_src}. They are however insufficient for accurate calculation of the luminosity and spectrum of unresolved sources and their removal. 

In order to proceed further we relied on the assumption that the combined emission of faint compact sources of all types has sufficiently  hard spectrum, harder than the emission of the ISM, and that the ISM makes the dominant contribution to the strong soft component observed in the majority of our galaxies (Fig.\ref{fig:diffuse_spectra}). This assumption is justified by observations of many individual objects and combined spectra of gas-free early type galaxies. We thus restricted our analysis  to the soft $0.5-2$ keV band and did not attempt to draw any quantitative conclusions regarding its properties above 2 keV. 
Based on the extrapolation of scaling relations obtained for various types of faint compact sources in the hard band, we conclude  that in the 0.5--2 keV band, the background- and HMXB-subtracted  unresolved emission  is a reasonable approximation to the emission of the diffuse  ISM.  Furthermore, along with the total passband luminosities, we will use spectral fitting results to explicitly separate the thermal emission from the power law component, presumably associated with emission of faint sources with harder spectra.

As a caveat, we note that if a diffuse gaseous component of a $\ga$(few) keV temperature is present in starburst galaxies, it would not be identified in the analysis  of \citet{2011MNRAS.418.1901B}, hence their scaling relation for young compact objects would overestimate  their luminosity and the presence of such hot ISM component would be missed in our analysis as well. Similarly, although the majority of faint compact sources do have rather hard spectra, there are stellar sources characterized by  sub-keV emission temperatures. They would be attributed to the ISM component in our analysis. Same is true about the supernova remnants -- with the temperature in the $\sim 0.5-1$ keV range \citep[e.g.][]{2010ApJS..187..495L} their spectra are somewhat harder than the soft ISM component (see below), but they will contribute to the overall unresolved emission in the 0.5--2 keV range.  These uncertainties appear to be currently unavoidable.

\section{Spectral analysis}
\label{sec:spectral_analysis}

We analyzed the background subtracted spectra in the 0.5--8 keV band. The spectra were approximated by a model consisting of one or, if required by the data, two thermal components  ({\tt mekal} in XSPEC) and a power law. The metallicity of the thermal component was fixed at the solar value. To this model we applied double photoelectric absorption component ({\tt phabs*phabs}), of which one was fixed at the Galactic value and the second was left as a free parameter in order to allow for any absorption intrinsic to the target galaxy. To this model we added a component accounting for the contribution of unresolved HMXBs, which parameters were fixed at the values determined as described in the Sect. \ref{sec:hmxb_removal}. Details of the best-fitting models for our sample galaxies are described in Table \ref{table:spectral_analysis}. 

In all galaxies, at least one thermal component was required, with plasma temperature in the range of $\sim 0.2 - 0.3$ keV. For about $\sim 1/3$  of galaxies the second thermal component was also statistically significantly  required by the data (F-test probability smaller than $10^{-3}$). Its temperature  was in the range of $\sim 0.5 - 0.9$ keV. These spectral characteristics are typical of the diffuse emission from normal starburst galaxies and consistent with measurements published by other authors for several galaxies from our sample \citep[e.g.][]{2004ApJS..151..193S, 2005ApJ...628..187G}. 

For $\sim$half of galaxies no intrinsic absorption was required by the data   (as before, we used the F-test probability of  $10^{-3}$ as the threshold). In these galaxies the intrinsic $N_{\rmn{H}}$ was fixed at zero value.

The power law component was included in order to account for the emission from unresolved compact sources, other than high-mass X-ray binaries, whose contribution could not be accurately determined from the mass and star formation rate of the galaxy  (see Sect.\ref{sec:contamination}). 
For the majority of galaxies the power law slope was poorly constrained, we therefore  fixed the slope to the value of $\Gamma = 1.7$ expected for low-mass X-ray binaries and let only its normalization be a free parameter of the fit. In four galaxies (NGC~2139, NGC~4194, NGC~5474 and NGC~7090), no power law component was required by the data. In these cases it was excluded from the fit. 

The luminosity of the best fit power law component in the soft and hard band for each galaxy is listed in the first two columns of the Table \ref{table:unres_src} along with the predicted luminosities of various unresolved components calculated in the section  \ref{sec:contamination}, and their sum. All luminosities are corrected for the Galactic and intrinsic absorption. As it is obvious from the Table, the sum of the predicted luminosities of LMXBs, CV/ABs and YSOs often exceed the observed luminosity of the power law component, in agreement with the considerations presented in the Section \ref{sec:contam_concl}.

\begin{figure*}
\begin{center}
\hbox
{
\includegraphics[width=0.5\linewidth]{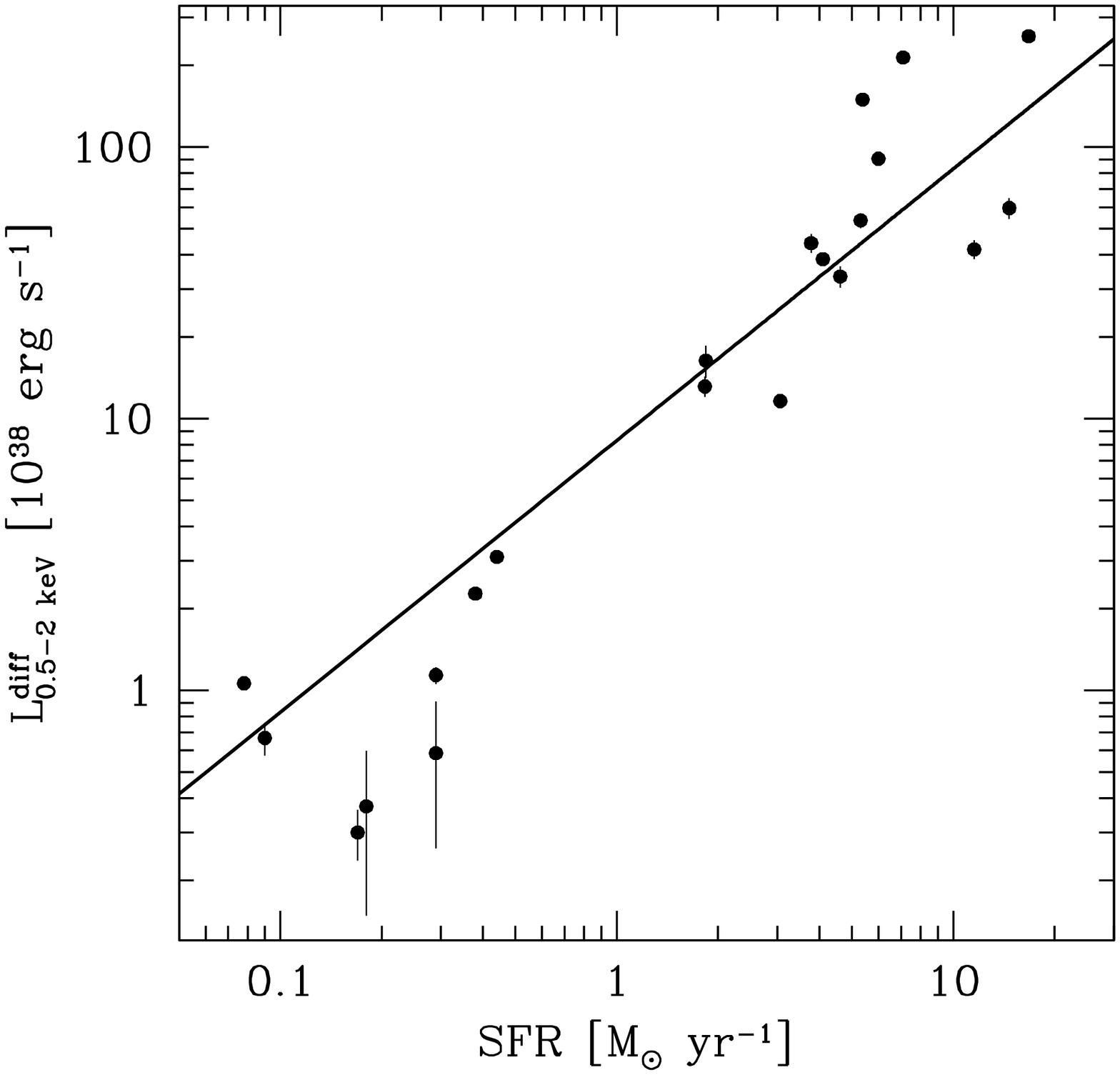}
\includegraphics[width=0.5\linewidth]{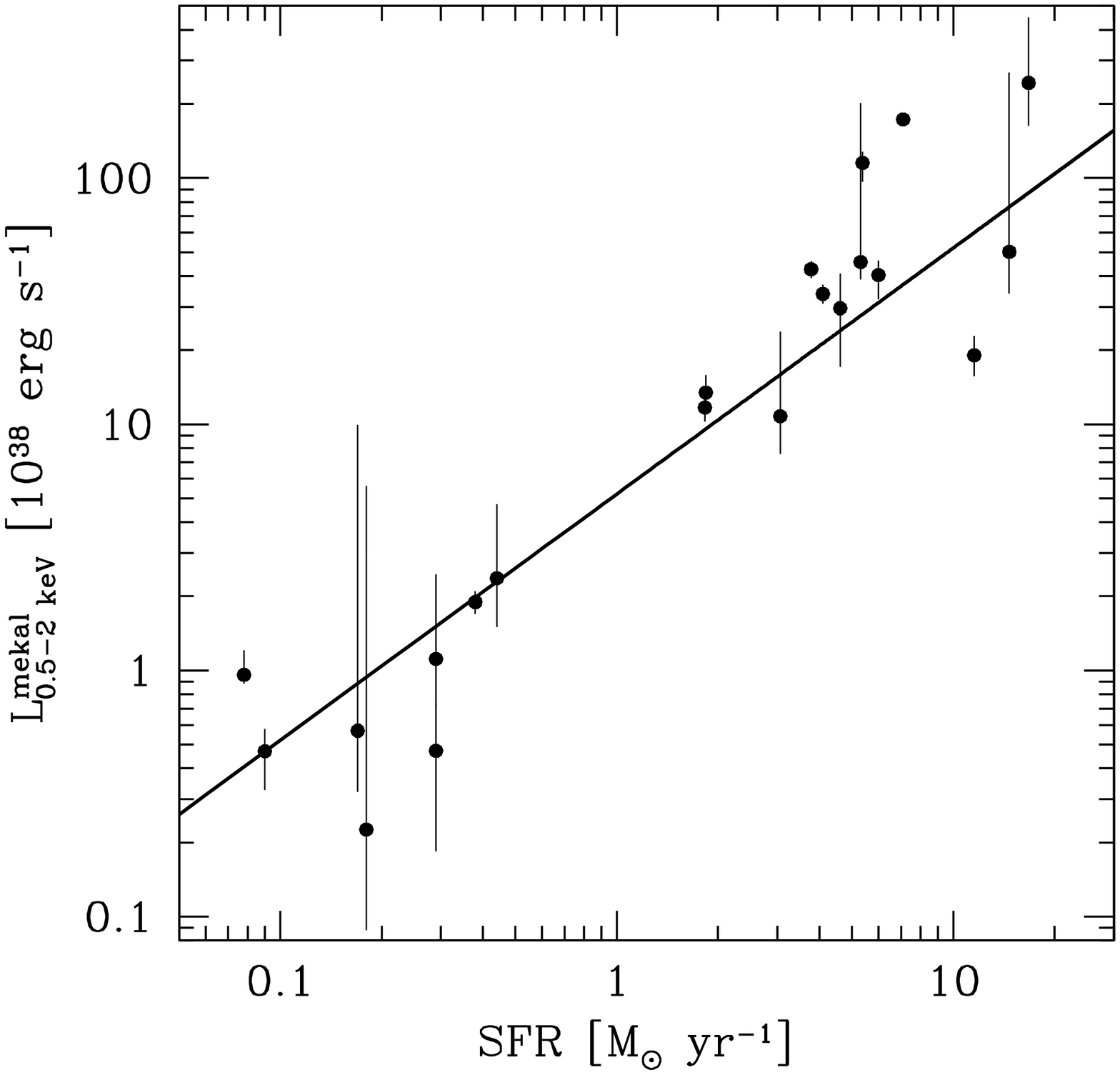}
}
\caption{The $L_{\rmn{X}}-\rmn{SFR}$ relation for the diffuse ISM emission in the 0.5--2 keV band. In the left panel, the ISM luminosity was computed as the total luminosity of the unresolved emission from which contribution of backgrounds and unresolved HMXBs was subtracted (see section \ref{sec:gas_luminosity}). The right hand panel shows the luminosity of the {\tt mekal} component(s). In both panels the luminosity is corrected for the Galactic absorption.} The best fit relations, described by equation (\ref{eq:ldiff_sfr}) and (\ref{eq:lmekal_sfr}), are shown by the solid line. The errors are given at $1\sigma$ confidence level.
\label{fig:lgas_sfr}
\end{center}
\end{figure*}

\section{X-ray luminosity of the diffuse ISM}
\label{sec:gas_luminosity}

We calculated luminosity of the diffuse ISM using a few different approximations,  and, finally, estimated its bolometric luminosity.

\subsection{The 0.5-2 keV luminosity}

First, we measured the 0.5--2 keV luminosity of the unresolved emission, from which contribution of backgrounds and unresolved HMXBs was subtracted. 
For each galaxy, we computed count-to-erg conversion for the best fit spectral model, which included correction for Galactic absorption. Using these conversion coefficients we calculated the energy flux and luminosity of the diffuse emission from the observed count rate of the unresolved emission. Before the conversion we subtracted from the latter the count rate of the backgrounds and the predicted count rate of unresolved HMXBs. We measured the luminosity in the 0.5--2 keV energy band and named it as diffuse emission luminosity $L^{\rmn{diff}}_{0.5-2\,\rmn{keV}}$. As discussed in the section \ref{sec:contamination}, the 2--8 keV band is likely subject to significant uncertainties due to contribution of various types of faint unresolved sources, we therefore do not attempt to perform any quantitative analysis in this energy band. The so obtained luminosities are listed in column 5 of Table \ref{table:spectral_analysis} and plotted against the SFR of the host galaxy in the left hand panel of Fig. \ref{fig:lgas_sfr}. The existence of a strong correlation is obvious from the plot. We fitted the data with a power law model $\log L^{\rmn{diff}}_{\rmn{X}} = \log A + \beta\,\log \rmn{SFR}$, using the $\chi^2$ minimization technique and obtained the best fit slope $\beta = 1.06$. The error on the slope computed with the standard $\Delta\chi^2=1$ prescription is $\sim 0.01$, thus, from the formal point of view, the best fit value of $\beta$ statistically significantly  deviates from unity. However, as the $\chi^2$ is very large in the minimum, $\chi^2\sim 6500$ for 19 d.o.f., the formal statistical error is not applicable. More appropriate is the value, computed via least square covariance matrix, which takes into account the scatter of the data points. The so computed error is  $\sigma_\beta\sim 0.1$.  With this, more realistic  error, the slope is compatible with unity. We therefore fit the $L_{\rmn{X}}-\rmn{SFR}$ data with the linear formula $L_{\rmn{X}}=A\times\,\rmn{SFR}$ using $\chi^2$ minimization technique and obtained the following relation: 
\begin{equation}
\label{eq:ldiff_sfr}
L^{\rmn{diff}}_{0.5-2\,\rmn{keV}} (\rmn{erg}\,\rmn{s}^{-1}) \approx (8.3\pm 0.1) \times 10^{38}\,\rmn{SFR}\,(M_{\odot}\,\rmn{yr}^{-1})
\end{equation}
with a dispersion of $\sigma = 0.34$  dex. The luminosity in the above relation is corrected for Galactic absorption. The error is $1\sigma$ for one parameter of interest and accounts for statistical uncertainties only. The advantage of this method is that it has minimal dependence on the details of the spectral model. 

As another approximation to the ISM luminosity we compute the 0.5--2 keV luminosity of the {\tt mekal} component, corrected for the Galactic absorption (sum of luminosities, when the second thermal component was required by the data). The so obtained luminosities are listed in column 6 of Table \ref{table:spectral_analysis} and plotted against the SFR of the host galaxy in the right-hand panel of Fig. \ref{fig:lgas_sfr}. The best-fit  slope for the $L_{\rmn{mekal}}-\rmn{SFR}$ relation is  $\beta = 0.94$. As the $\chi^2$ is still rather large in the minimum ($\chi^2=504.4$ for 19 dof), we rely on the  error computed from the covariance matrix,  $\sigma_\beta\sim 0.1$ and conclude that  the slope is consistent with unity.
Fixing the slope at unity, we obtain the following linear relation:
\begin{equation}
\label{eq:lmekal_sfr}
L^{\rmn{mekal}}_{0.5-2\,\rmn{keV}} (\rmn{erg}\,\rmn{s}^{-1}) \approx (5.2\pm 0.2) \times 10^{38}\,\rmn{SFR}\,(M_{\odot}\,\rmn{yr}^{-1})
\end{equation}
with a dispersion of $\sigma = 0.34$ dex. In computing the $\chi^2$ we approximately  took into account the asymmetry of  the luminosity errors by using the positive or negative uncertainty depending on the sign of the deviation of the data from the model.

Comparing equations (\ref{eq:ldiff_sfr}) and (\ref{eq:lmekal_sfr}) we can estimate that the average contribution of sources with hard spectra to the 0.5--2 keV luminosity of the unresolved emission is $\approx 37\%$. However, values for individual galaxies show considerable scatter with the $rms\sim 22\%$, i.e. about the half of this value. A fraction of this hard emission is  associated with various types of unresolved faint compact sources, but some fraction can, in principle, be due to the contribution of the hotter gas, unaccounted by the low temperature {\tt mekal} components.

Finally, we note that the result of the fitting of the $L_{\rmn{X}}-\rmn{SFR}$ relation depends not only on whether the least square of $\chi^2$ minimization is used, but also on whether the fitting is performed in linear or logarithmic space. For example, the least squares fit to the data in the log space,  $\log L_{\rmn{X}} = \log A + \log \rmn{SFR}$, gives close but not same value of the scale factor in eq. (\ref{eq:ldiff_sfr}), $7.1\cdot 10^{38}$ erg/s per $M_\odot$/yr ($5.7\cdot 10^{38}$ in eq. (\ref{eq:lmekal_sfr})). The difference is larger for the $\chi^2$ minimization in the logarithmic space ($16.2\cdot 10^{38}$ and $11.6\cdot 10^{38}$ respectively). This should be taken into account in comparing results of different analyses.

\subsection{Bolometric luminosity of gas}
\label{sec:lbol}

In many tasks (e.g. energy balance of the ISM or calculation of the supernova feedback) the knowledge of the bolometric luminosity of ISM is critical. However its calculation is impeded by a number of factors, all related to the smallness of the gas temperature. Below we attempt to compute the scaling relation for the  bolometric luminosity of ISM. We do it in several steps, with each step  producing better approximation to intrinsic bolometric luminosity of gas, but at the same time obtaining progressively more model dependent result. 

For the XSPEC's {\tt mekal} model with  temperature in the range $kT=0.24-0.71$, the bolometric correction for the 0.5--2 keV luminosity ranges from $\approx 3$ to $\approx 1.4$. For the two temperature spectrum with the emission measures equal to the average emission measures from the Table \ref{table:spectral_analysis}, the bolometric correction factor is $\approx 2$. Applying this correction to the eq.(\ref{eq:lmekal_sfr}) we obtain the scale factor for the {\em apparent} bolometric luminosity: $L_{\rmn{bol}}/\rmn{SFR}\sim 10^{39}$ erg/s per $M_\odot$/yr. This relation is corrected for the Galactic absorption but does not account for possible intrinsic absorption in galaxies.

For almost a half of galaxies a model with intrinsic absorption gave a statistically significant improvement of the fit (Table \ref{table:spectral_analysis}). The obtained values of $N_{\rmn{H}}$ are in the $\sim (1-4)\cdot 10^{21}$ cm$^{-2}$ range, i.e. in the range of values expected for star-forming galaxies.  For the majority of remaining galaxies, the upper limits for the intrinsic absorption are of the same order. It is not clear, whether intrinsic absorption is real or is an artifact of an inadequate spectral model. 
Even if it is real and measured with reasonable accuracy in our spectral fits, the calculation of the intrinsic luminosity of the gas involves large absorption correction factors, due to small temperature and relatively large $N_{\rmn{H}}$. The final result is sensitive to the details of the spectral model at low energies, below the low energy threshold of Chandra detectors. It is further complicated by the degeneracy between the model parameters resulting in rather large uncertainties in the computed luminosity.

\begin{figure}
\begin{center}
\includegraphics[width=1.0\linewidth]{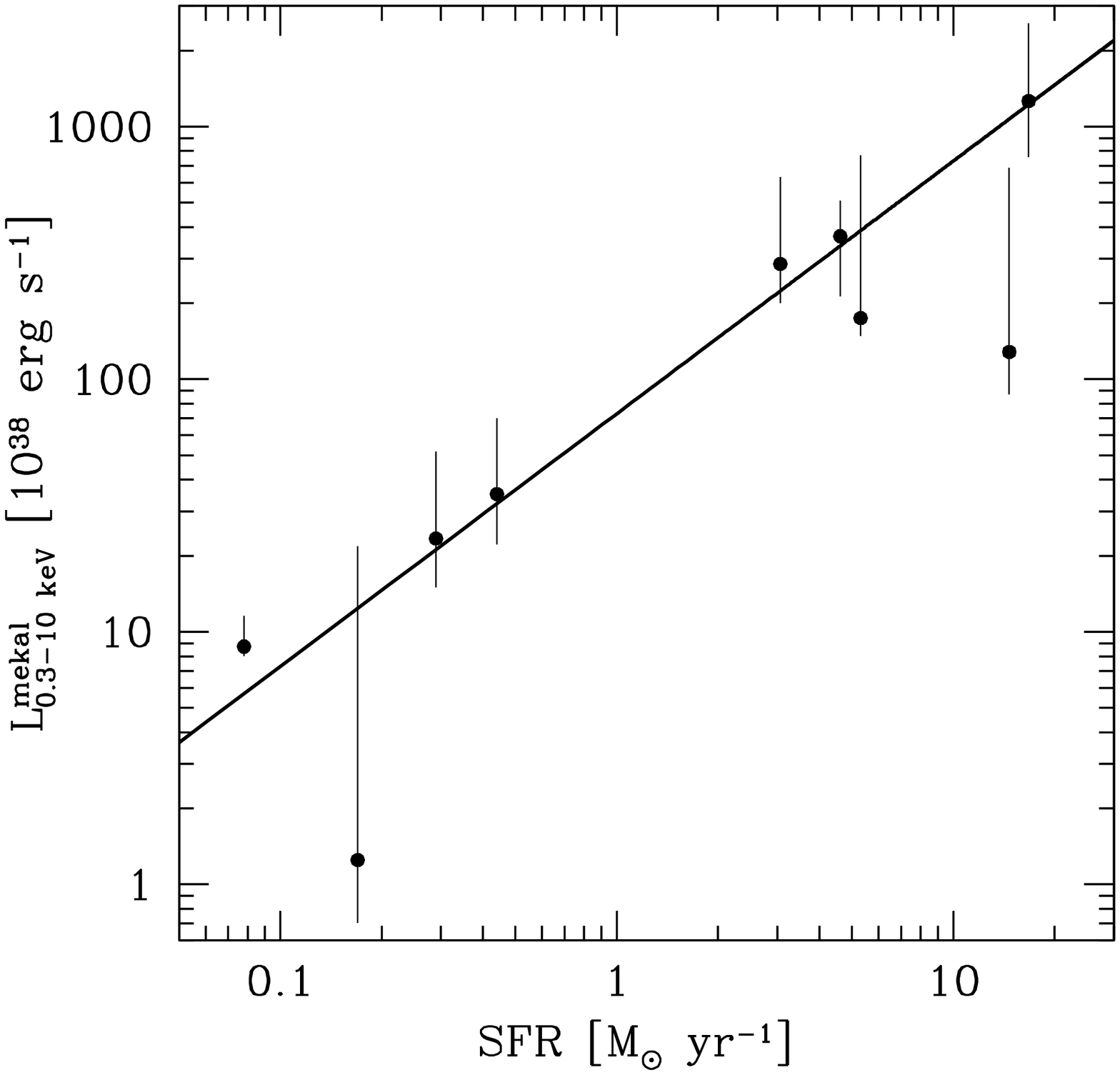}
\caption{The $L_{\rmn{X}}-\rmn{SFR}$ relation for the 0.3--10 keV luminosity of the thermal component(s), corrected for both the Galactic and intrinsic absorption. The best fit relation, described by equation (\ref{eq:lbol_sfr}), are shown by the solid line. The errors are given at $1\sigma$ confidence level.} 
\label{fig:lbol_sfr}
\end{center}
\end{figure}

With these caveats in mind, we attempt to derive the scaling relation for the {\em intrinsic} bolometric luminosity of the ISM.
For galaxies for which intrinsic absorption was required by the data, we compute the 0.3--10 keV luminosity of thermal component(s) corrected for both Galactic and intrinsic absorption. The so obtained luminosities are listed in column 7 of Table \ref{table:spectral_analysis} and plotted in  Fig. \ref{fig:lbol_sfr}.
The $\chi^2$ fit to the data with the  linear relation gives:
\begin{equation}
\label{eq:lbol_sfr}
L^{\rmn{mekal}}_{0.3-10\,\rmn{keV}} (\rmn{erg}\,\rmn{s}^{-1}) = (7.3\pm 1.3) \times 10^{39}\,\rmn{SFR}\,(M_{\odot}\,\rmn{yr}^{-1})
\end{equation}
As before, we took into account the asymmetry of the error bars for the luminosity, which in this case notably affected the result, due to strongly asymmetric errors for some of the galaxies (Fig.\ref{fig:lbol_sfr}).  In computing the best fit, we excluded NGC 1569 (the leftmost point in  Fig. \ref{fig:lbol_sfr}). This galaxy has the largest value of $L_{\rmn{X}}/\rmn{SFR}$ and the smallest value of the negative uncertainty, significantly smaller than other galaxies in our sample, and strongly affects the best fit, shifting it in the  upward direction by a factor of 1.5. 

Applying the bolometric correction for the average spectrum, $L_{\rmn{bol}}/L_{0.3-10\,\rmn{keV}}\approx 2$, calculated above, we estimate  the scaling relation for the {\em intrinsic} bolometric luminosity of gas: $L_{\rmn{bol}}/\rmn{SFR}\sim 1.5\cdot 10^{40}$  erg/s per $M_\odot$/yr. Comparing this with the apparent bolometric luminosity, derived in the beginning of this subsection, we conclude that correction for intrinsic absorption introduces  a factor of $\sim 10$. As discussed above, due to several unavoidable uncertainties these numbers should be treated with certain caution.

\begin{figure*}
\begin{center}
\hbox
{
\includegraphics[width=0.5\linewidth]{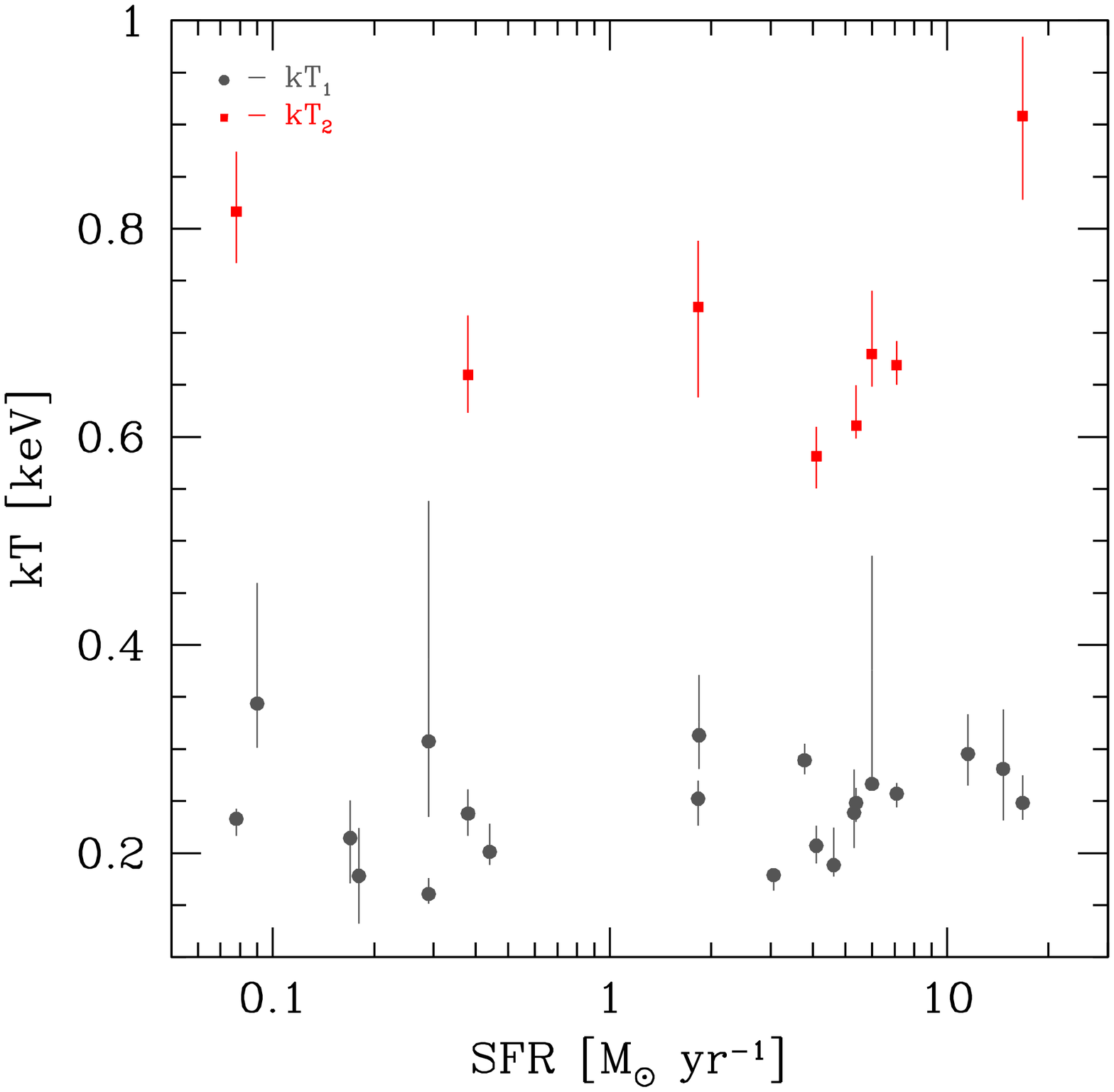}
\includegraphics[width=0.5\linewidth]{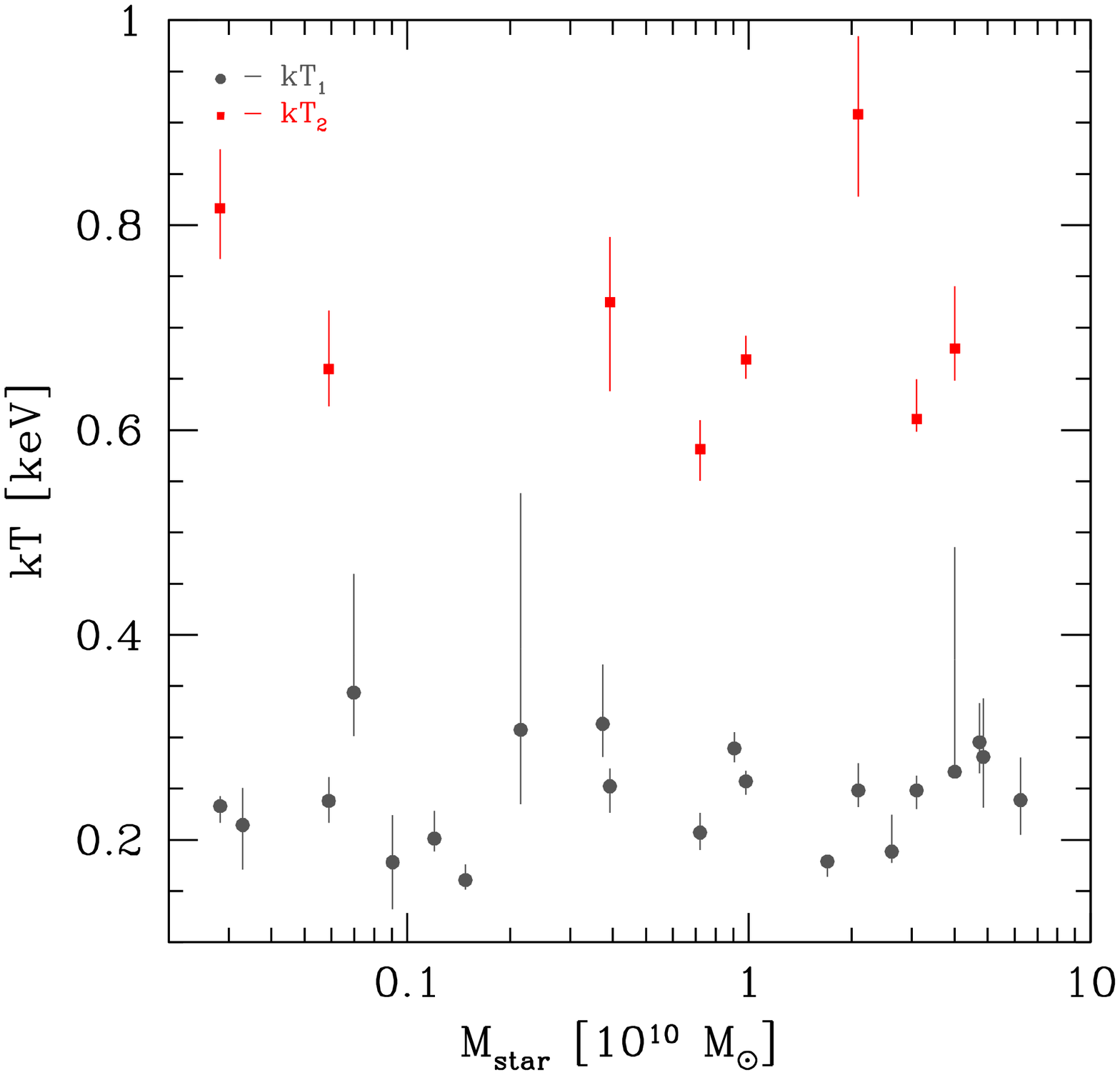}
}
\caption{Gas temperature of the thermal components for the X-ray diffuse emission spectra plotted against the SFR (left) and the stellar mass (right) of the galaxy. Filled circles (grey) and squares show  temperatures of the cooler and hotter component respectively. Note that the latter was not required in all galaxies. The errors are given at $1\sigma$ confidence level.}
\label{fig:kT_sfr_mstar}
\end{center}
\end{figure*}

\section{Discussion}

\subsection{Comparison with previous results}

Based on XMM-{\it Newton} observations of a sample of nearby late-type galaxies, \citet{2009MNRAS.394.1741O} studied the unresolved emission in the 0.3--1 keV band. They measured the soft $L_{\rmn{X}}/\rmn{SFR}$ ratio in the range from $2.5\times 10^{38}$ to $\sim 10^{39}$ ($\rmn{erg}\,\rmn{s}^{-1})/(M_{\odot}\,\rmn{yr}^{-1})$. In their analysis, they removed the contribution of the most luminous point sources only, therefore the unresolved emission may include some contributions of X-ray binaries. Their SFR determination was based on FUV luminosities and corrected for dust attenuation effects. For our sample of resolved galaxies there is a close correspondence between the attenuation corrected UV-based SFR and the total SFR estimator adopted in the present work (see Fig. 12 of Paper I). Therefore, to compare our results with those of \citet{2009MNRAS.394.1741O}, we only need to convert the X-ray luminosities from $0.3-1.0$ to $0.5-2.0$ keV band. To do this conversion, we assumed a two component spectrum consisting of the optically thin plasma component with temperature $kT=0.24$ keV and a power law component with the photon index $\Gamma=2$, with the absorption column density of $N_{\rmn{H}}=3\times 10^{21}$ cm$^{-2}$. The ratio between the normalizations of the power law and thermal component was fixed at $1.36\times 10^{-2}$, the average value determined from the spectral fits. This led to the conversion factor of  0.86. With this value of the conversion factor, their $L_{\rmn{X}}/\rmn{SFR}$ ratios lie in the range  from $2.9\cdot 10^{38}$ to $\sim 1.2\cdot10^{39}$ ($\rmn{erg}\,\rmn{s}^{-1})/(M_{\odot}\,\rmn{yr}^{-1})$, in reasonable agreement with the scale factors given by our eqs. (\ref{eq:ldiff_sfr}) and ({\ref{eq:lmekal_sfr}). Furthermore,  to be closer to the derivation  of  \citet{2009MNRAS.394.1741O}, we did not subtract  contribution of unresolved HMXBs and obtained a scaling factor of $7.5\cdot 10^{38}$ ($\rmn{erg}\,\rmn{s}^{-1})/(M_{\odot}\,\rmn{yr}^{-1}$), in yet better agreement with their range of $L_{\rmn{X}}/\rmn{SFR}$ values. 

\citet{2005ApJ...628..187G} computed the $L_{0.3-2\,\rmn{keV}} /L_{\rmn{FIR}}$ ratios for a sample of ultra-luminous infrared starburst and dwarf starburst galaxies and derived a mean $L_{0.3-2\,\rmn{keV}} /L_{\rmn{FIR}}$ ratio of $10^{-4}$ with a  dispersion of $\sim 0.4$ dex. Their energy band,   $0.3-2.0$ keV, is  almost identical  to the $0.5-2.0$ keV band used in this work. In order to compare our results with the derivations from these authors, we only converted the FIR ($42.5-122.5\,\mu$m) luminosity to total IR ($8-1000\,\mu$m) band using a conversion factor of 1.7. The latter was then converted into SFR using the value of $L_{IR}/\rmn{SFR}=2.2\cdot 10^{43} (\rmn{erg}\,\rmn{s}^{-1})/(M_{\odot}\,\rmn{yr}^{-1})$ from \citet{1998ARA&A..36..189K}. We thus found $L_{\rmn{X}}/\rmn{SFR}\approx 1.3\cdot 10^{39}$ ($\rmn{erg}\,\rmn{s}^{-1})/(M_{\odot}\,\rmn{yr}^{-1})$. This value exceeds by a factor of $\sim 2$  our eqs (\ref{eq:ldiff_sfr}) and  (\ref{eq:lmekal_sfr}) and is also near the high end of the $L_{\rmn{X}}/\rmn{SFR}$ determined by  \citet{2009MNRAS.394.1741O}. On the other hand, it is $\sim 5$ times smaller than our estimate of the absorption corrected  0.3--10 keV luminosity, eq.(\ref{eq:lbol_sfr}). We note that  \citet{2005ApJ...628..187G} did not do any removal of point sources, they rather relied on the spectral separation of the contributions of gas and compact sources. 

\citet{2012arXiv1201.0551L} recently computed the relation between the X-ray luminosity of the galactic coronae and the SFR for a large sample of nearby edge-on galaxies. For the absorption corrected 0.5--2 keV luminosity of the thermal emission, they found $L_{\rmn{X}}/\rmn{SFR}\approx 1.3\cdot 10^{39}$ ($\rmn{erg}\,\rmn{s}^{-1})/(M_{\odot}\,\rmn{yr}^{-1})$, larger than our eq.(\ref{eq:ldiff_sfr}) and  (\ref{eq:lmekal_sfr}), but  smaller than our estimate of the 0.3--10 keV luminosity, eq.(\ref{eq:lbol_sfr}). In estimating the X-ray luminosity, they removed the detected point sources and accounted for the contribution of unresolved CVs and ABs in the spectral fits. 

Thus, there is a scatter of a factor of up to $\sim2-5$ between the scale factors obtained by different authors.
This is not surprising, given the difference in the calculation methods, bright compact source removal and absorption correction or lack of thereof, in the  galactic samples and sizes of the analyzed regions between different authors.

\subsection{Physical parameters of the gas}

We plot in the Fig.\ref{fig:kT_sfr_mstar} the best fit gas temperatures vs SFR and the stellar mass of the galaxy. 
As it is obvious from the plot, there are statistically significant variations in the gas temperature between galaxies. These variations have  $rms=0.05$ keV and $rms=0.11$ keV for the low-temperature and high-temperature plasma respectively. However, we did not find any obvious trends with the SFR and stellar mass of the galaxy.

Knowing the emission measure and the volume occupied by gas  we can estimate the gas density. Assuming that gas occupies a disk like volume with thickness of the order of the exponential scale height of the ISM determined from its X-ray emission \citep{2012arXiv1201.0551L} we find the gas densities in the $\sim 10^{-3}-10^{-2}$ cm$^{-3}$ range. The radiative  cooling times of the gas are in the $\sim 10^7-10^9$ years range. The high end of this range, $\sim 10^9$ yrs, found for some of the galaxies, is likely to  be caused by too simplified assumptions about the shape of  the volume occupied by the hot plasma.

\subsection{Efficiency of the ISM heating by supernovae}
\label{sec:phys_prop}

The existence of the linear relation between the ISM luminosity and the star formation rate allows one to estimate the efficiency of the ISM heating by the supernovae, assuming that they are the main source of energy powering the ISM emission \citep[e.g.][]{1985Natur.317...44C, 2000AJ....120.2965S, 2004ApJ...606..829S, 2005ApJ...628..187G, 2008MNRAS.388...56B}. Indeed, for the majority of galaxies the gas cooling times are relatively short, suggesting that an external energy input is required in order to maintain the X-ray luminosity of the gas throughout the star formation event. Considering only core collapse supernovae, their rate can be related to the SFR: $\sim 1$ SN per 100 yrs per 1 $M_\odot$/yr \citep{2012A&A...537A.132B}. Assuming further that one supernova releases $E_{SN}=10^{51}$ erg of mechanical energy, the rate of the mechanical energy release due to core-collapse supernovae is: $\dot{E}_{SN}\approx 3.2\cdot 10^{41}\times \rmn{SFR}$ erg/s. Comparing it with the relation for the {\em intrinsic} bolometric luminosity of the gas given in the end of the Section \ref{sec:gas_luminosity}, we constrain the efficiency of the conversion of the mechanical energy of the core collapse supernovae in to thermal energy of ISM: 
\begin{equation}
\label{eq:sn_eff}
\epsilon_{SN}\sim 5\cdot 10^{-2} \left(\frac{E_{SN}}{10^{51}}\right)^{-1}
\end{equation}
Of course all caveat regarding the accuracy of the estimation of the bolometric luminosity of ISM given in  the Section 6,  are fully applicable here. For example, using the scaling relation for the {\em apparent} bolometric luminosity, $L_{\rmn{X}} \sim 1 \cdot 10^{39}\times \rmn{SFR}$, one would obtain an order of magnitude smaller number, $\epsilon_{SN} \sim 3\cdot 10^{-3}$.

Eq. (\ref{eq:sn_eff}) differs by a factor of $\sim$few from recent results by \citet{2012arXiv1201.0551L}, who found the energy conversion efficiency of core collapse supernovae in the range $2.0\times 10^{-3}$ to $1.2\times 10^{-2}$. We believe that the main source of discrepancy is the fact that  \citet{2012arXiv1201.0551L} used the {\em apparent} luminosity of ISM, uncorrected for possible  intrinsic absorption in star-forming galaxies, whereas the eq.(\ref{eq:sn_eff}) relies on our estimate of the {\em intrinsic} bolometric luminosity of the gas.

\subsection{Contribution of the ISM emission to the total X-ray luminosity due to star formation}

Comparing the relations eq.(\ref{eq:ldiff_sfr})-(\ref{eq:lbol_sfr}) with the results of Paper I we conclude that above 0.5 keV,   the apparent luminosity of hot ISM is about  $\approx 20\%$ of the HMXB luminosity. The behavior of intrinsic luminosities depends on the details of the spectrum of the collective emission of HMXBs at low energies and significance of the intrinsic absorption. With present numbers, assuming  that there is no significant soft component in the HMXB spectra, the intrinsic X-ray luminosity of the ISM may be comparable or even exceed that of HMXBs.

\section{Conclusions}
For the sample of nearby star-forming galaxies from Paper I, we studied characteristics of the diffuse emission. We took special care of various systematic effect and we believe that we are in good control of the contamination of unresolved emission by bright compact sources of all types and by unresolved faint high-mass X-ray binaries. The contribution of other types of faint sources (LMXBs, CVs and ABs and young stellar objects) can not be accurately subtracted given the present knowledge of their scaling relations. We argued that their contribution in the soft band is not dominant. 

At least one soft thermal component is statistically significantly required in all galaxies in our sample. Its has the mean  temperature of $\left<kT\right>\approx 0.24$ keV with $rms=0.05$ keV. In about $\sim 1/3$ of galaxies,  the second thermal component is required by the data, with the mean temperature of $\left<kT\right>\approx 0.71$ keV and $rms=0.11$ keV. Although we observe statistically significant differences in temperature between galaxies, no clear trends with the stellar mass or star formation rate could be found. 

The luminosity of diffuse emission in the $0.5-2$ keV band, as well as luminosity of its thermal component(s), correlate well with the star formation rate. The corresponding scaling relations are described by eqs.(\ref{eq:ldiff_sfr})-(\ref{eq:lbol_sfr}). We attempt to estimate the bolometric luminosity of the gas and obtained results differing by an order of magnitude, $L_{\rmn{X}}/\rmn{SFR}\sim 10^{39}-10^{40}  (\rmn{erg}\,\rmn{s}^{-1})/(M_{\odot}\,\rmn{yr}^{-1})$, depending on whether the intrinsic absorption was allowed for. Our theoretically most accurate, but in practice the most model dependent result for the {\em intrinsic} bolometric luminosity is $L_{\rmn{bol}}/\rmn{SFR}\sim 1.5\cdot 10^{40} (\rmn{erg}\,\rmn{s}^{-1})/(M_{\odot}\,\rmn{yr}^{-1})$. This number exceeds by about an order of magnitude the scaling relations obtained previously \citep{2005ApJ...628..187G, 2009MNRAS.394.1741O, 2012arXiv1201.0551L}. Various caveats associated with its derivation are discussed in the section \ref{sec:lbol}. Using this scaling relation, we estimate the efficiency of supernova feedback into thermal energy of ISM, $\epsilon_{SN}\sim 5\%$.

\section*{Acknowledgments}
SM gratefully acknowledges financial support through the NASA grant AR1-12008X. The authors thank the anonymous referee for helpful comments that improved this paper.
This research made use of \textit{Chandra} archival data and software provided by the \textit{Chandra} X-ray Center (CXC) in the application package CIAO. This research has made use of SAOImage DS9, developed by Smithsonian Astrophysical Observatory.
The \textit{Spitzer Space Telescope} is operated by the Jet Propulsion Laboratory, California Institute of Technology, under contract with the NASA. \textit{GALEX} is a NASA Small Explorer, launched in 2003 April.
This publication makes use of data products from Two Micron All Sky Survey, which is a joint project of the University of Massachusetts and the Infrared Processing and Analysis Center/California Institute of Technology, funded by the NASA and the National Science Foundation.
This research has made use of the NASA/IPAC Extragalactic Database (NED) which is operated by the Jet Propulsion Laboratory, California Institute of Technology, under contract with the National Aeronautics and Space Administration.

\bsp

\begin{landscape}
\begin{table}
\begin{minipage}{220mm}
\centering
\caption{The observed luminosity of the power law component of the unresolved emission and predicted contributions of various types of faint sources in soft and hard bands.}
\label{table:unres_src}
\begin{tabular}{@{}l l l l c l c l c l c@{}}
\hline
& \multicolumn{2}{|c|}{{\sc power law}\,\footnote{Soft and hard band luminosities of the power law component determined from the spectral fitting of the spectrum of the unresolved emission from which backgrounds and contribution of unresolved HMXBs (Table \ref{table:unres_hmxb}) was subtracted (sect. \ref{sec:spectral_analysis}), corrected for both Galactic and intrinsic absorptions.}} & \multicolumn{2}{|c|}{{\sc unresolved CVs and ABs}\,\footnote{Predicted luminosities of unresolved CVs and ABs (Sect. \ref{sec:lmxb_removal}).}} & \multicolumn{2}{|c|}{{\sc unresolved LMXBs}\,\footnote{Predicted luminosities of unresolved LMXBs (Sect. \ref{sec:lmxb_removal}).}} & \multicolumn{2}{|c|}{{\sc unresolved YSOs}\,\footnote{Predicted luminosities of unresolved YSOs (Sect. \ref{sec:yso_removal}).}}& \multicolumn{2}{|c|}{{\sc total unresolved}\footnote{Sum of the predicted luminosities of LMXBs, CV/ABs and YSOs.}}\\
Galaxy & $L_{0.5-2\,\rmn{keV}}$ & $L_{2-8\,\rmn{keV}}$ &\vline\, $L_{0.5-2\,\rmn{keV}}$ & $L_{2-8\,\rmn{keV}}$ &\vline\, $L_{0.5-2\,\rmn{keV}}$ & $L_{2-8\,\rmn{keV}}$ &\vline\, $L_{0.5-2\,\rmn{keV}}$ & $L_{2-8\,\rmn{keV}}$ &\vline\, $L_{0.5-2\,\rmn{keV}}$ & $L_{2-8\,\rmn{keV}}$\\
 & $(\rmn{erg}\,\rmn{s}^{-1})$ & $(\rmn{erg}\,\rmn{s}^{-1})$ & \vline\, $(\rmn{erg}\,\rmn{s}^{-1})$ & $(\rmn{erg}\,\rmn{s}^{-1})$ & \vline\, $(\rmn{erg}\,\rmn{s}^{-1})$ & $(\rmn{erg}\,\rmn{s}^{-1})$ & \vline\, $(\rmn{erg}\,\rmn{s}^{-1})$ & $(\rmn{erg}\,\rmn{s}^{-1})$ & \vline\, $(\rmn{erg}\,\rmn{s}^{-1})$ & $(\rmn{erg}\,\rmn{s}^{-1})$\\	
\hline
\hline
NGC 0278       & $3.39\times10^{38}$ & $5.14\times10^{38}$ &\vline\, $4.85\times10^{37}$ & $4.85\times10^{37}$ &\vline\, $2.15\times10^{38}$ & $2.84\times10^{38}$ &\vline\, $7.48\times10^{38}$ & $6.68\times10^{38}$ &\vline\, $1.01\times10^{39}$ & $1.00\times10^{39}$\\ 
NGC 0520       & $1.80\times10^{39}$ & $2.72\times10^{39}$ &\vline\, $1.87\times10^{38}$ & $1.86\times10^{38}$ &\vline\, $1.38\times10^{39}$ & $1.81\times10^{39}$ &\vline\, $2.11\times10^{39}$ & $1.88\times10^{39}$ &\vline\, $3.68\times 10^{39}$ & $3.88\times 10^{39}$\\ 
NGC 1313       & $1.08\times10^{38}$ & $1.63\times10^{38}$ &\vline\, $7.85\times10^{36}$ & $7.84\times10^{36}$ &\vline\, $4.00\times10^{36}$ & $5.28\times10^{36}$ &\vline\, $8.03\times10^{37}$ & $7.17\times10^{37}$ &\vline\, $9.21\times 10^{37}$ & $8.48\times 10^{37}$\\ 
NGC 1569       & $1.42\times10^{37}$ & $2.15\times10^{37}$ &\vline\, $1.19\times10^{36}$ & $1.19\times10^{36}$ &\vline\, $1.20\times10^{35}$ & $1.59\times10^{35}$ &\vline\, $1.42\times10^{37}$ & $1.27\times10^{37}$ &\vline\, $1.55\times 10^{37}$ & $1.40\times 10^{37}$\\ 
NGC 2139       & $-$ & $-$ &\vline\, $7.39\times10^{37}$ & $7.38\times10^{37}$ &\vline\, $2.57\times10^{38}$ & $3.39\times10^{38}$ &\vline\, $6.90\times10^{38}$ & $6.16\times10^{38}$ &\vline\, $1.02\times 10^{39}$ & $1.03\times 10^{39}$\\ 
NGC 3079       & $4.42\times10^{39}$ & $6.70\times10^{39}$ &\vline\, $1.95\times10^{38}$ & $1.95\times10^{38}$ &\vline\, $1.23\times10^{39}$ & $1.62\times10^{39}$ &\vline\, $1.09\times10^{39}$ & $9.76\times10^{38}$ &\vline\, $2.52\times 10^{39}$ & $2.79\times 10^{39}$\\ 
NGC 3310       & $3.70\times10^{39}$ & $5.60\times10^{39}$ &\vline\, $8.14\times10^{37}$ & $8.14\times10^{37}$ &\vline\, $2.59\times10^{38}$ & $3.41\times10^{38}$ &\vline\, $1.29\times10^{39}$ & $1.16\times10^{39}$ &\vline\, $1.63\times 10^{39}$ & $1.58\times 10^{39}$\\ 
NGC 3556       & $5.53\times10^{37}$ & $8.38\times10^{37}$ &\vline\, $9.57\times10^{37}$ & $9.56\times10^{37}$ &\vline\, $1.92\times10^{38}$ & $2.53\times10^{38}$ &\vline\, $5.58\times10^{38}$ & $4.99\times10^{38}$ &\vline\, $8.46\times 10^{38}$ & $8.48\times 10^{38}$\\ 
NGC 3631       & $1.01\times10^{39}$ & $1.54\times10^{39}$ &\vline\, $1.48\times10^{38}$ & $1.48\times10^{38}$ &\vline\, $6.09\times10^{38}$ & $8.03\times10^{38}$ &\vline\, $8.41\times10^{38}$ & $7.52\times10^{38}$ &\vline\, $1.60\times 10^{39}$ & $1.70\times 10^{39}$\\ 
NGC 4038/39 & $3.31\times10^{39}$ & $5.01\times10^{39}$ &\vline\, $1.42\times10^{38}$ & $1.42\times10^{38}$ &\vline\, $8.18\times10^{38}$ & $1.08\times10^{39}$ &\vline\, $9.82\times10^{38}$ & $8.77\times10^{38}$ &\vline\, $1.94\times 10^{39}$ & $2.10\times 10^{39}$\\ 
NGC 4194       & $-$ & $-$ &\vline\, $1.27\times10^{38}$ & $1.27\times10^{38}$ &\vline\, $9.21\times10^{38}$ & $1.21\times10^{39}$ &\vline\, $3.06\times10^{39}$ & $2.73\times10^{39}$ &\vline\, $4.11\times 10^{39}$ & $4.07\times 10^{39}$\\ 
NGC 4214       & $1.91\times10^{36}$ & $2.90\times10^{36}$ &\vline\, $2.21\times10^{36}$ & $2.21\times10^{36}$ &\vline\, $2.42\times10^{36}$ & $3.19\times10^{36}$ &\vline\, $3.10\times10^{37}$ & $2.77\times10^{37}$ &\vline\, $3.56\times 10^{37}$ & $3.31\times 10^{37}$\\ 
NGC 4490       & $1.65\times10^{38}$ & $2.50\times10^{38}$ &\vline\, $3.37\times10^{37}$ & $3.37\times10^{37}$ &\vline\, $1.00\times10^{38}$ & $1.32\times10^{38}$ &\vline\, $3.34\times10^{38}$ & $2.98\times10^{38}$ &\vline\, $4.68\times 10^{38}$ & $4.64\times 10^{38}$\\ 
NGC 4625       & $9.86\times10^{36}$ & $1.49\times10^{37}$ &\vline\, $3.94\times10^{36}$ & $3.93\times10^{36}$ &\vline\, $4.91\times10^{36}$ & $6.47\times10^{36}$ &\vline\, $1.64\times10^{37}$ & $1.47\times10^{37}$ &\vline\, $2.53\times 10^{37}$ & $2.51\times 10^{37}$\\ 
NGC 5253       & $3.37\times10^{37}$ & $5.11\times10^{37}$ &\vline\, $4.48\times10^{36}$ & $4.47\times10^{36}$ &\vline\, $5.27\times10^{36}$ & $6.94\times10^{36}$ &\vline\, $6.93\times10^{37}$ & $6.19\times10^{37}$ &\vline\, $7.91\times 10^{37}$ & $7.33\times 10^{37}$\\ 
NGC 5474       & $-$ & $-$ &\vline\, $5.85\times10^{36}$ & $5.85\times10^{36}$ &\vline\, $4.77\times10^{36}$ & $6.29\times10^{36}$ &\vline\, $3.28\times10^{37}$ & $2.93\times10^{37}$ &\vline\, $4.34\times 10^{37}$ & $4.14\times 10^{37}$\\ 
NGC 5775       & $5.43\times10^{38}$ & $8.22\times10^{38}$ &\vline\, $3.24\times10^{38}$ & $3.23\times10^{38}$ &\vline\, $1.60\times10^{39}$ & $2.11\times10^{39}$ &\vline\, $9.69\times10^{38}$ & $8.66\times10^{38}$ &\vline\, $2.89\times 10^{39}$ & $3.30\times 10^{39}$\\ 
NGC 7090       & $-$ & $-$ &\vline\, $1.53\times10^{37}$ & $1.53\times10^{37}$ &\vline\, $4.44\times10^{37}$ & $5.85\times10^{37}$ &\vline\, $5.29\times10^{37}$ & $4.73\times10^{37}$ &\vline\, $1.13\times 10^{38}$ & $1.21\times 10^{38}$\\ 
NGC 7541       & $5.67\times10^{38}$ & $8.59\times10^{38}$ &\vline\, $3.13\times10^{38}$ & $3.13\times10^{38}$ &\vline\, $1.83\times10^{39}$ & $2.41\times10^{39}$ &\vline\, $2.68\times10^{39}$ & $2.39\times10^{39}$ &\vline\, $4.82\times 10^{39}$ & $5.11\times 10^{39}$\\ 
NGC 7793       & $1.60\times10^{37}$ & $2.43\times10^{37}$ &\vline\, $8.98\times10^{36}$ & $8.97\times10^{36}$ &\vline\, $1.82\times10^{37}$ & $2.40\times10^{37}$ &\vline\, $5.29\times10^{37}$ & $4.73\times10^{37}$ &\vline\, $8.01\times 10^{37}$ & $8.03\times 10^{37}$\\ 
UGC 05720     & $2.49\times10^{37}$ & $3.77\times10^{37}$ & \vline\, $2.10\times10^{37}$ & $2.10\times10^{37}$ &\vline\, $1.56\times10^{38}$ & $2.05\times10^{38}$ &\vline\, $3.36\times10^{38}$ & $3.00\times10^{38}$ &\vline\, $5.13\times 10^{38}$ & $5.26\times 10^{38}$\\ 
\hline
\end{tabular}
\end{minipage}
\end{table}
\end{landscape}

\begin{landscape}
\begin{table}
\begin{minipage}{230mm}
\centering
\caption{X-ray spectral analysis of the diffuse emission.}
\label{table:spectral_analysis}
\begin{tabular}{@{}l c c c c c c c c c c c c@{}}
\hline
\vspace{1mm}
Galaxy & SFR\footnote{Star formation rate from Spitzer and GALEX data (see Sect. 6 Paper I).}  &  $M_{\star}$\footnote{Stellar mass from 2MASS data (see Sect. 5 Paper I).} & $i$\footnote{Inclination angles taken from \citet{1988Sci...242..310T}.} & $\log L_{\rmn{diff}}$\footnote{Luminosity of the diffuse emission in the 0.5--2 keV band corrected for Galactic absorption, backgrounds and contribution of unresolved HMXBs subtracted (Sect. \ref{sec:gas_luminosity}).} & $\log L_{\rmn{mek}}$\footnote{Luminosity of the thermal component of unresolved emission in the $0.5-2\,\rmn{keV}$ band, corrected for Galactic absorption (Sect. \ref{sec:gas_luminosity}).} & $\log L^{\rmn{corr}}_{\rmn{mek}}$\footnote{Luminosity of the thermal component of unresolved emission in the $0.3-10\,\rmn{keV}$ band corrected for Galactic and intrinsic absorption (Sect. \ref{sec:gas_luminosity}).} & $N_{\rmn{H,int}}$\footnote{Intrinsic hydrogen column density, errors are quoted at 90\% confidence level.} & $kT_{1}$\footnote{Low-temperature plasma component of the spectral model, errors are quoted at 90\% confidence level.} & $\left(n_e^2V\right)_1$\footnote{The emission measure of the first thermal component, errors are quoted at 90\% confidence level.} & $kT_{2}$\footnote{High-temperature plasma component of the spectral model, errors are quoted at 90\% confidence level.}  & $\left(n_e^2V\right)_2$\footnote{The emission measure of the second thermal component, errors are quoted at 90\% confidence level.}  & $\chi^{2}/\nu$\footnote{$\chi^{2}$ divided by the number of degrees of freedom for the best fit spectral model.}  \\
& ($M_{\odot}/\rmn{yr}$) &  ($10^{10}\,M_{\odot}$) & (deg) & $(\rmn{erg}\,\rmn{s}^{-1})$ & $(\rmn{erg}\,\rmn{s}^{-1})$ & $(\rmn{erg}\,\rmn{s}^{-1})$ & ($10^{22}\rmn{cm}^{-2}$) & (keV) & ($10^{-3}\rmn{cm}^{-6}\,\rmn{kpc}^{3}$)  & (keV) & ($10^{-3}\rmn{cm}^{-6}\,\rmn{kpc}^{3}$) &  \\ 
\hline
\hline
NGC~0278	& 4.1 & 0.7					& 0& 39.59 & 39.53 & - & - & 0.21$_{-0.03}^{+0.03}$ & 3.46$^{+0.63}_{-0.65}$ & 0.58$_{-0.05}^{+0.05}$ & 2.58$^{+0.57}_{-0.52}$ & 98.21/99 \\
NGC~0520 	& 11.6 & 4.7 	  				& 69 & 39.62 & 39.28 & - & - & 0.30$_{-0.05}^{+0.06}$ & 3.42$^{+1.11}_{-0.99}$ & - & - & 153.60/173 \\
NGC~1313 	& 0.44 & 0.1 	  				& 38 & 38.49 & 38.38 & 39.55 & 0.41$_{-0.06}^{+0.09}$ & 0.20$_{-0.02}^{+0.04}$ & 6.50$^{+10.63}_{-3.92}$ & - & - & 129.29/108\\
NGC~1569 	&  $7.8\times10^{-2}$ 	& $2.8\times10^{-2}$ 	& 64 & 38.03 & 37.98 & 38.94 & 0.42 $_{-0.05}^{+0.05}$ & 0.23$_{-0.03}^{+0.02}$ & 1.28$^{+0.77}_{-0.19}$ & 0.82$_{-0.08}^{+0.09}$ & 0.14$^{+0.04}_{-0.04}$ & 239.49/174\\
NGC~2139 	& 3.8 & 0.91 		  			& 53 & 39.65 & 39.63 & - & - & 0.29$_{-0.02}^{+0.03}$ & 7.76$^{+1.01}_{-0.99}$ & - & - & 112.52/113\\
NGC~3079 	& 6.0 & 4.0 					& 88 & 39.96 & 39.61 & - & - &  0.27$_{-0.27}^{+0.20}$ & 1.02$^{+1.80}_{-0.80}$ & 0.68$_{-0.05}^{+0.10}$ & 5.16$^{+0.88}_{-1.84}$ & 108.87/108 \\
NGC~3310 	& 7.1 & 0.98 					& 25 & 40.33 & 40.24 & - & - & 0.26$_{-0.02}^{+0.02}$ & 15.93$^{+1.08}_{-2.14}$ & 0.67$_{-0.03}^{+0.04}$ & 12.48$^{+1.77}_{-1.77}$ & 248.17/190 \\
NGC~3556 	& 3.1 & 1.7 					& 81 & 39.06 & 39.03 & 40.46 & 0.46$_{-0.06}^{+0.09}$ & 0.18$_{-0.02}^{+0.01}$ & 59.42$^{+117.75}_{-29.10}$ & - & -  & 247.60/226 \\
NGC~3631 	& 4.6 & 2.6 					& 36 & 39.52 & 39.47 & 40.57 & 0.35$_{-0.18}^{+0.10}$ & 0.19$_{-0.02}^{+0.06}$ & 72.82$^{+45.45}_{-50.34}$ & - & - & 478.31/492  \\
NGC~4038/39 	& 5.4 & 3.1 					& - & 40.17 & 40.06 & - & - & 0.25$_{-0.03}^{+0.02}$ & 7.86$^{+3.80}_{-1.27}$ & 0.61$_{-0.02}^{+0.06}$ & 10.31$^{+1.16}_{-4.05}$ & 490.88/468 \\
NGC~4194 	& 16.8 & 2.1 					& 49 & 40.41 & 40.39 & 41.10 & 0.31$_{-0.11}^{+0.16}$ & 0.25$_{-0.03}^{+0.04}$ & 172.11$^{+349.40}_{-134.75}$ & 0.91$_{-0.13}^{+0.12}$ & 30.07$^{+12.23}_{-6.50}$ & 98.86/112 \\
NGC~4214 	& 0.17 & $3.3\times10^{-2}$ 			& 37 & 37.48 & 37.59 & 38.10 & 0.14$_{-0.14}^{+0.34}$ & 0.21$_{-0.07}^{+0.06}$ & 0.22$^{+5.87}_{-0.16}$ & - & -  & 165.19/170 \\
NGC~4490 	& 1.8 & 0.39 					& 65 & 39.12 & 39.07 & - & - & 0.25$_{-0.04}^{+0.03}$ & 1.37$^{+0.17}_{-0.42}$ & 0.72$_{-0.14}^{+0.10}$ & 0.70$^{+0.29}_{-0.15}$ & 290.44/293 \\
NGC~4625 	& 0.09 & $7.0\times10^{-2}$ 			& 23 & 37.82 & 37.67 & - & - & 0.34$_{-0.07}^{+0.19}$ & 0.07$^{+0.03}_{-0.03}$ & - & - & 44.55/37 \\
NGC~5253 	& 0.38 & $5.9\times10^{-2}$ 			& 77 & 38.36 & 38.28 & - & - & 0.24$_{-0.03}^{+0.04}$ & 0.20$^{+0.05}_{-0.04}$ & 0.66$_{-0.06}^{+0.09}$ & 0.13$^{+0.03}_{-0.03}$ & 172.36/200 \\
NGC~5474 	& 0.18 & $9.1\times10^{-2}$ 			& 37 & 37.57 & 37.35 & - & - & 0.18$_{-0.08}^{+0.08}$ & 0.07$^{+2.73}_{-0.07}$ & - & - & 104.63/118  \\
NGC~5775 	& 5.3 & 6.3 					& 84 & 39.73 & 39.66 & 40.24 & 0.20$_{-0.09}^{+0.23}$ & 0.24$_{-0.06}^{+0.07}$ & 28.96$^{+162.76}_{-7.09}$ & - & -  & 113.74/135 \\
NGC~7090 	& 0.29 & 0.22 					& 90 & 37.77 & 37.67 & - & - & 0.31$_{-0.12}^{+0.38}$ & 0.08$^{+0.08}_{-0.08}$ & - & - & 111.66/119 \\
NGC~7541 	& 14.7 & 4.9 					& 75 & 39.77 & 39.70 & 40.11 & 0.15$_{-0.11}^{+0.28}$& 0.28$_{-0.08}^{+0.09}$ & 19.97$^{+142.69}_{-10.57}$ & - & - & 66.00/63 \\
NGC~7793 	& 0.29 & 0.15					& 50 & 38.06 &	38.05 & 39.37 & 0.38$_{-0.17}^{+0.10}$& 0.16$_{-0.02}^{+0.02}$ & 5.62$^{+11.13}_{-3.30}$ & - & - & 163.08/176  \\
UGC~05720 	& 1.8 & 0.37 					& 28 & 39.21 & 39.13 & - & - & 0.31$_{-0.05}^{+0.09}$ & 	2.42$^{+0.70}_{-0.62}$ & - & -  & 11.75/15 \\                                             
\hline 
\end{tabular}
\end{minipage}
\end{table}
\end{landscape}

\label{lastpage}

\end{document}